\documentclass[reprint,groupedaddress,amsmath,amssymb,aip,prl,floatfix]{revtex4-1}
\usepackage[nolist,nohyperlinks]{acronym}
\usepackage{xcolor} 
\usepackage[utf8]{inputenc} 
\usepackage{graphicx}
\usepackage{dcolumn}
\usepackage{upgreek}
\usepackage{bm}
\usepackage{braket}
\renewcommand{\thefigure}{\arabic{figure}}
\renewcommand{\thetable}{\arabic{table}}
\newacro{cQED}{circuit quantum electrodynamics}
\newacro{VNA}{vector network analyzer}
\newacro{HEMT}{high electron mobility transistor}
\newacro{SNR}{signal-to-noise ratio}
\newcommand{\Q}[1]{Q_{\text{#1}}}
\newcommand{\particip}[1]{p_{#1}} 

\begin{document}

\preprint{APS/123-QED}

\title{A coaxial line architecture for integrating and scaling 3D cQED systems}

\author{C. Axline}
\author{M. Reagor}
\author{R. Heeres}
\author{P. Reinhold}
\author{C. Wang}
\author{K. Shain}
\author{W. Pfaff}
\author{Y. Chu}
\author{L. Frunzio}
\author{R. J. Schoelkopf}
\affiliation{Department of Applied Physics, Yale University, New Haven, Connecticut 06511, USA}

\date{\today}

\begin{abstract}
Numerous loss mechanisms can limit coherence and scalability of planar and 3D-based \ac{cQED} devices, particularly due to their packaging. The low loss and natural isolation of 3D enclosures make them good candidates for coherent scaling. We introduce a coaxial transmission line device architecture with coherence similar to traditional 3D \ac{cQED} systems. Measurements demonstrate well-controlled external and on-chip couplings, a spectrum absent of cross-talk or spurious modes, and excellent resonator and qubit lifetimes. We integrate a resonator-qubit system in this architecture with a seamless 3D cavity, and separately pattern a qubit, readout resonator, Purcell filter and high-$Q$ stripline resonator on a single chip. Device coherence and its ease of integration make this a promising tool for complex experiments. 
\end{abstract}

\maketitle


\section{Introduction} 

Engineered quantum systems are becoming increasingly complex, with single packages incorporating on the order of ten coherent elements (resonators and qubits) used to store or process quantum information\cite{barends_superconducting_2014}. Tens of logical modules, or hundreds of elements, are needed to build systems capable of quantum error correction and operations with logical qubits\cite{steane_overhead_2003}. As they scale, these larger systems must retain the level of control and coherence of smaller systems in order to achieve scalable levels of performance.

Multiple approaches are being explored to build scalable systems using superconducting microwave circuits. Each design must grow in size and complexity to include many elements, expanding to incorporate both planar and 3D structures\cite{gambetta_building_2015,brecht_multilayer_2016}. The ultimate size of scaled systems, regardless of architecture, will likely be set by enclosure hardware and connector volume. Even fields generated from planar structures are three-dimensional, so the importance of package and interconnect design cannot be dismissed if existing systems are to be reliably expanded.

Complex circuits have highlighted some of these packaging challenges. In lithographically-patterned circuits, cross-talk and package modes can be more difficult to suppress given the presence of wirebonds, circuit board materials, isolated ground planes, and connectors\cite{chen_fabrication_2014}. Although they contain fewer loss-inducing elements, multi-cavity 3D waveguide designs may be limited by seam dissipation\cite{brecht_demonstration_2015, kirchmair_observation_2013}. Enclosure and connector quality become more important in scaled systems of all types. Reduced coherence from these effects can inhibit expansion in its current form. 

We aim to create a platform for system expansion in which many lithographically-defined elements can be coherently combined. To incorporate the longest-lived resources available to \ac{cQED}, we would also like to add 3D cavities in a lossless manner. This coherent complexity could pave the way for error-correctable modules\cite{devoret_superconducting_2013,nickerson_freely_2014}. 

Cavity resonators constitute a good platform from which to reach this goal. Here, the cavity and the enclosure are one and the same, providing a well-controlled spectrum of modes. Transmon qubits in 3D cavities have shown exceptionally high coherence\cite{paik_observation_2011,dial_bulk_2015}. These cavities can be made seamless to limit dissipation while remaining integrable\cite{reagor_quantum_2015}. These features form the basis for a low-loss enclosure suited to expansion.

Here we demonstrate a carefully engineered 3D waveguide package for \ac{cQED} devices. By enclosing chip-based circuit elements in such a package, the resulting coaxial transmission line (``coax-line'') device can be highly coherent and extendable. Addressing a host of likely losses within the package, including coupling, seams, and materials, we produce chip-based element single photon relaxation rates at the level of the state of the art ($\sim 50$\,$\upmu$s). Resonators, qubits, and filters are fabricated on a single chip. We find that coupling between these elements can be well-controlled by changes in circuit parameters set with lithographic precision. Finally, we integrate this system with millisecond 3D cavities and show that the combined systems remain highly coherent. In the near-term, this platform allows for significantly more complex many-resonator, many-qubit circuits. When combined with more advanced techniques for fabricating 3D enclosures using lithography and multi-wafer bonding\cite{brecht_multilayer_2016}, the coax-line provides an attractive route towards long-term scaling.

\section{Design and Performance}
\begin{figure}[tbp]
  \centering
  \includegraphics[width=0.8\linewidth]{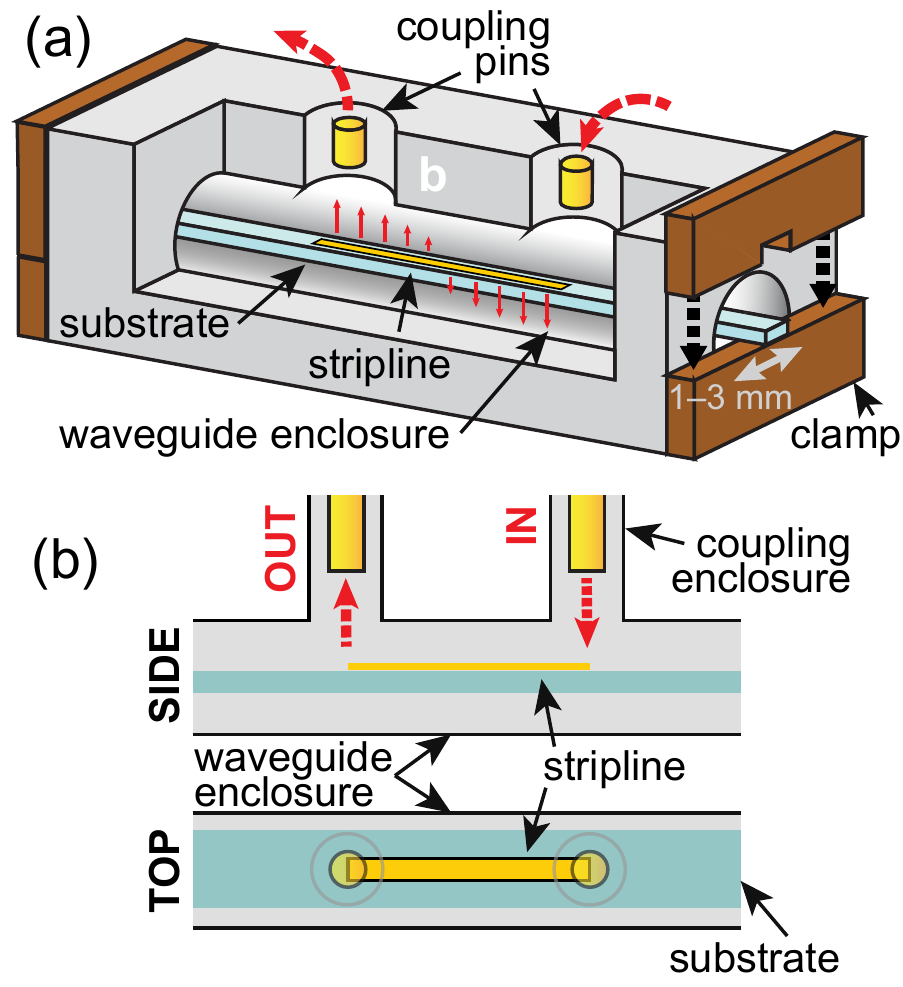}
  \caption{(a) A depiction of the coax-line architecture includes a patterned sapphire chip (blue) inserted into a tubular enclosure and clamped at both ends (brown). Indium wire (not shown) secures the chip within the clamps. The resonator is patterned in the center of the chip, acting as the center conductor of a coaxial transmission line resonator. The enclosure ends are many attenuation lengths away from the resonator. Small red arrows represent the electric field pattern of the $\lambda/2$ resonant mode. Dashed arrows indicate the input and output paths used in transmission measurements. Coupling pins (gold), recessed within two smaller waveguides that intersect the primary enclosure, couple evanescently and carry signals to external connectors. (b) The configuration of couplers in a symmetric transmission experiment.}
     \label{design}
\end{figure}

We design and measure circuits placed in 3D enclosures that avoid many common forms of loss in \ac{cQED} systems. A seamless circular waveguide forms the package enclosure and acts as a ground plane (Figure~\ref{design}a). Circuit elements are patterned on a sapphire substrate to define each mode of the device. Deposited and machined metals are both chosen to be aluminum. Where no metal is present on-chip, the waveguide attenuates signals below its cutoff frequency (typically 40\,GHz). The chip is suspended within the enclosure by clamps at each end, where the fields from critical circuit elements are exponentially attenuated.

We evaluate device performance beginning with one simple element: a resonator. Choosing a quasi-stripline architecture, we pattern a $\lambda/2$ resonator on the substrate and position it near the center of the enclosure. The resonant frequency is primarily determined by the length of the conducting strip, but also depends on chip size, chip placement, and enclosure diameter.

Inspired by the robust coupling method used in 3D cavities, we introduce input and output signals using two evanescently-coupled pins within sub-cutoff waveguides that intersect the primary waveguide enclosure. Pins are recessed to an adjustable depth within each coupling enclosure, located above each end of the stripline. Both pins are used in transmission measurements (Figure~\ref{design}b). Just one pin can be used to measure in reflection or feedline-coupled transmission\cite{_apl_????}. As described later, this approach yields predictable couplings that can be varied over a wide dynamic range without compromising package integrity.

\begin{figure*}[tbp]
  \centering
  \includegraphics[width=0.95\linewidth]{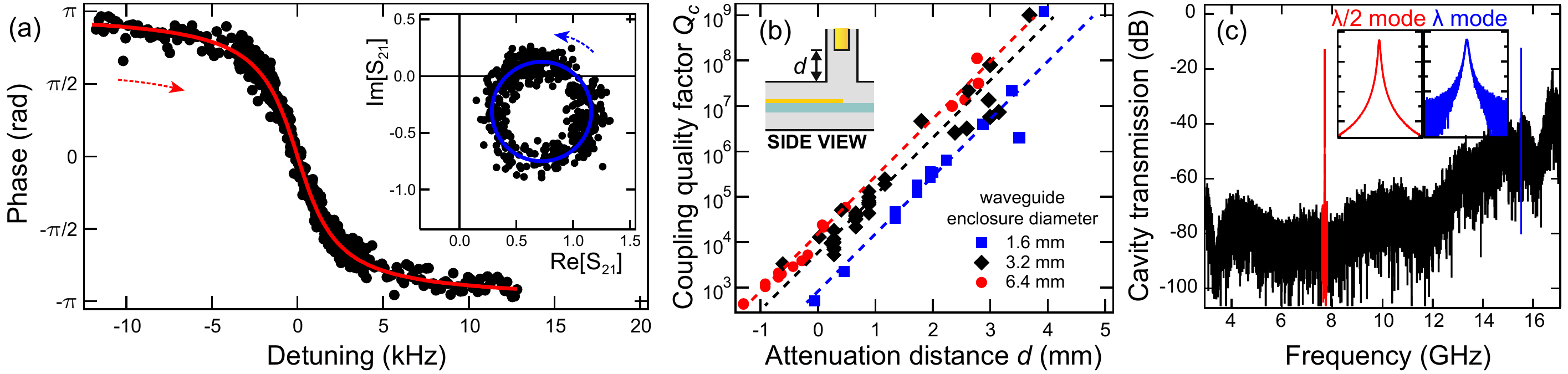}
  \caption{(a) Resonator quality factors $\Q{i}$ and $\Q{c}$ are extracted from fits of the resonance circle in $S_{21}$ (inset) and the phase response (main figure)\cite{khalil_analysis_2012}. Data are measured at an average cavity photon number $\bar{n} \approx 1$ where unsaturated defects produce 30--50\% lower $Q$ relative to higher powers\cite{gao_experimental_2008}. A representative sample, plotted, has $\Q{i}=5.98\pm0.07$~million and $\Q{c}=4.27\pm0.07$~million. Dashed arrows indicate frequency sweep direction. (b) Coupling quality factor $\Q{c}$ is measured for 12\,mm-long resonators in enclosures with three different diameters (solid points). The attenuation distance $d$ (inset) is measured from the end of the coupling pin to the edge of the enclosure, and enters the enclosure for $d<0$. We measure $\Q{c}$ as high as $10^9$, above which reduced \ac{SNR} hinders measurement for our typical $\Q{i}$'s. Measured $\Q{c}$'s follow exponential behavior (dashed lines) with minor deviations due to resonator shape, chip placement, and machining variance. This suggests that no unknown mode couples more strongly than $\Q{c}=10^9$. (c) Transmission measurements show only the expected resonant modes, here at 7.7\,GHz and 15.5\,GHz (insets have 300\,kHz span). The fundamental mode sees isolation of $>60$\,dB. The noise floor is due to the frequency dependence of the readout system and \ac{HEMT} noise.}
     \label{spurious}
\end{figure*}

To verify that these devices are suited to complex experiments, we first demonstrate long lifetimes. For resonators, this requires achieving a high internal quality factor ($\Q{i}$) at sufficiently weak coupling (high coupling quality factor $\Q{c}$). We measure these parameters by cooling coax-line resonators to $\sim$ 20\,mK and exciting single-photon-or-less circulating power. The devices are connected in a feedline-coupled configuration and transmission coefficient $S_{21}$ is measured using a \ac{VNA}. Coupling parameters are extracted from fits to $S_{21}$ (Figure~\ref{spurious}a inset). Measurements are usually performed in an undercoupled configuration ($\Q{c} \geq \Q{i}$, with $\Q{c}$ up to $10^9$) so that the total quality factor $Q$ ($1/Q = 1/\Q{i} + 1/\Q{c}$) is a direct measure of the internal losses. The best reported $\Q{i}$ for lithographically-defined aluminum-on-sapphire resonators fabricated under similar conditions (e-beam evaporation, no substrate annealing) is $\sim$1~million\cite{megrant_planar_2012}. We observe $\Q{i}$ as high as $(8.0 \pm 0.5) \times 10^{6}$ at single-photon power, surpassing this by about an order of magnitude. This suggests that quality factors in lithographic devices are not solely dependent on materials or fabrication methods, but are also affected by package contributions.

By restricting wave propagation to seamless waveguides with cutoffs far above the operating frequency, we demonstrate that mode coupling can be made arbitrarily weak without additional structures or filtering\cite{sandberg_radiation-suppressed_2013}. By varying the coupling attenuation distance and measuring $\Q{c}$, we see good agreement with the expected exponential scaling over six decades with no observed upper limit (Figure~\ref{spurious}b). Control over a large dynamic range in coupling strengths is possible by simply modifying pin length. Therefore, we can achieve very strong coupling ($\Q{c} \sim 10^{3}$) to some elements used for measurement or readout, at the same time as weak couplings ($\Q{c} \sim 10^{8}$) used to excite and control long-lived memory elements.

Another critical requirement of a properly-designed package is to prevent spurious electromagnetic modes. Using stronger coupling and a symmetric transmission configuration, we measured $S_{21}$ to determine the spectral ``cleanliness'' over a large range. Because the enclosure should attenuate any modes below its cutoff frequency in the absence of package seams, we expect the measured background to be low, dominated by the noise of other elements in the measurement chain. Figure~\ref{spurious}c shows a calibrated $S_{21}$ trace within the measurement bandwidth of our \ac{HEMT} amplifier. Aside from the fundamental ($\lambda/2$) and first harmonic ($\lambda$) modes arising from the stripline resonator, no other modes are observed. This confirms that good mode control can be achieved using a coax-line architecture. 

\begin{figure}[tbp]
  \centering
  \includegraphics[width=1.0\linewidth]{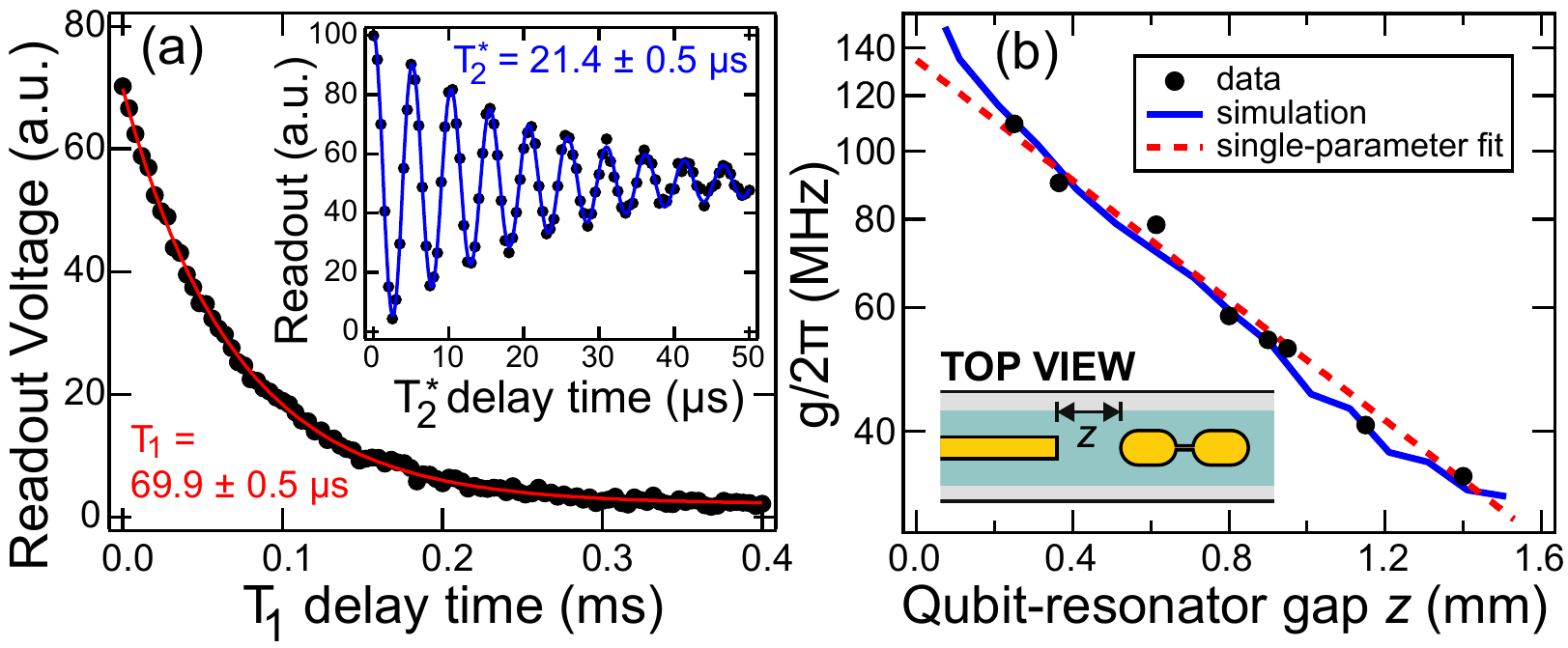}
  \caption{(a) Qubit $T_{1}$ (main figure) and $T_{2}^{*}$ (inset) of one characteristic device. The $T_{1}$ experiment is fit to an exponential (red), while the detuned $T_{2}^{*}$ Ramsey experiment is fit to an exponentially decaying sine function (blue). (b) Coupling between the qubit and stripline resonator is controlled by adjusting their end-to-end separation $z$ (inset). Values of effective qubit-stripline coupling rate $g$ are measured for different $z$ (black points) and fit using an exponential function $A e^{-\alpha z}$ (dashed red line) with single free parameter $A$. The calculated attenuation, 8.5\,dB/mm, comes from simulation of a 2.8\,mm-diameter waveguide with bare substrate. A full system finite-element simulation (solid blue line) predicts similar scaling.}
     \label{highQ}
\end{figure}

As the next step in increasing complexity, we pattern a transmon qubit alongside the resonator\cite{devoret_circuit-qed:_2007}. We characterize the system's coherence and compare measured parameters to simulation. We control the qubit using a weakly-coupled port and read out through the resonator and its more strongly coupled port. These qubits exhibit 30--80\,$\upmu$s lifetimes, near to the state of the art values for transmon $T_1$'s (Figure~\ref{highQ}a). This is equivalent to quality factors $\leq 3$ million, not far from those of resonators. Undercoupled resonators had equally high $\Q{i}$ with and without qubits. Important system parameters, such as mode frequencies and qubit anharmonicity, were found to agree well with predictions from finite-element simulations of the design\cite{nigg_black-box_2012, _apl_????}. This is the first indication that additional complexity can be added to a coax-line device without decreasing control over parameters or coherence values.

The presence of a lithographic resonator-qubit system enables us to test whether on-chip element coupling follows the same waveguide-attenuated behavior as external coupling. We expect that by varying the distance $z$ between element ends (Figure~\ref{highQ}b inset), the chip enclosure will exponentially attenuate electric field $|\vec{E}| \propto e^{-\alpha z}$. The resonator-qubit dispersive shift $\chi$ should scale as $\chi \sim |\vec{E}|^{2}$. However, different resonator-qubit detunings $\Delta$ between samples make direct comparison difficult. To relate them consistently to $z$, we calculate an effective coupling $g$ defined by the relation $\chi = 2 g^2/\Delta$, related to the detuned Jaynes-Cummings model\cite{wallraff_strong_2004}. When $z$ is varied experimentally, we find that the measured change in $g$ is consistent with a calculated waveguide attenuation scale length $1/\alpha \approx 1.02$\,mm, as well as with simulations (Figure~\ref{highQ}b). This suggests that no unintended coupling is present, and that reasonably small separations between elements can produce a range of qubit-resonator couplings useful for typical \ac{cQED} applications. 

The lifetimes of resonators and qubits in this system can be understood by examining the spatial participation of each mode in dissipative dielectrics and conductors. The large resonator mode volume dilutes lossy material participation---the same effect that increases coherence of 3D cavities relative to traditional planar circuits. By measuring resonators in waveguide enclosures with different diameters, we find that higher $\Q{i}$ generally corresponds with larger diameter\cite{_apl_????}. This scaling behavior is consistent with loss originating from waveguide surface resistance, a waveguide dielectric layer, or on-chip dielectric layers, but does not distinguish between these mechanisms.

Even though the enclosure body is seam-free, we can evaluate whether seams at each end introduce dissipation. We predict their effect using a model of seam loss as a distributed admittance\cite{brecht_demonstration_2015} applied to simulation. The simulation places a conservative bound, $\Q{i} \geq 10^{8}$, on typical designs separated 7\,mm from the waveguide end. Positioning striplines $\sim$ 3\,mm from one end of an enclosure produced an immeasurable effect on $\Q{i}$, thereby raising this bound. Since typical devices see significantly greater isolation from end seams, they appear unlikely to affect performance. Therefore, coax-line devices are well-positioned to act as a testbed for alternative loss mechanisms. Further studies will be required to pinpoint the dominant sources of loss, but the coherence levels already achieved allow us to increase the system's complexity further.

\section{Integration and Expansion}

\begin{figure}[tbp]
  \centering
  \includegraphics[width=0.95\linewidth]{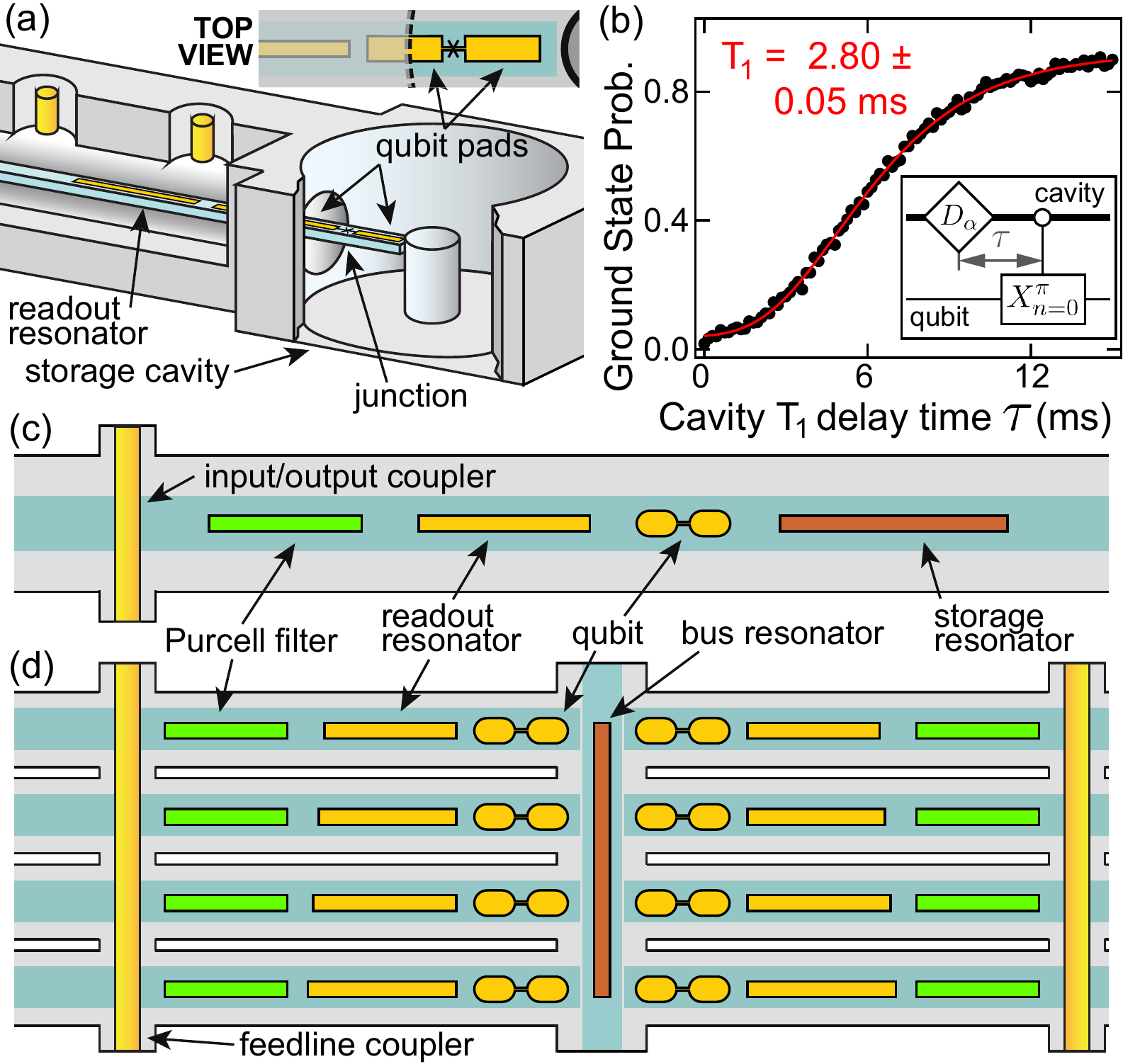}
  \caption{(a) Combining a chip-based circuit with a 3D coaxial stub cavity. The transmon qubit antenna pads straddle the circular waveguide enclosure and stub cavity. Qubit-stripline coupling is controlled lithographically, and qubit-cavity coupling is set by antenna geometry and chip position. (b) Cavity $T_{1}$ is fit (red) to data (black points). The cavity lifetime is not spoiled by the qubit's shorter lifetime. (c) We extended the chip-based qubit-stripline system by adding a Purcell filter and high-$Q$ stripline storage resonator. It requires a single pin-coupler in a feedline-coupled configuration. (d) We propose an expansion of multiple chip-based modules, in which eight Purcell-filtered qubits interact with a bus resonator and are addressed by two multiplexed readout lines.} 
     \label{integration}
\end{figure}

Many circuit elements must be integrated within a single enclosure to allow more versatile, hardware-efficient \ac{cQED} experiments. To demonstrate an instance of a long-lived element in the presence of significant complexity, we combine a very high-$Q$ 3D cavity with the coax-line architecture. In the resulting package (Figure~\ref{integration}a), the pads of a transmon qubit bridge two structures: the coax-line qubit-and-stripline system and a 3D coaxial stub cavity\cite{reagor_quantum_2015}.

We characterize parameters of the complete system, including coupling and coherence values. Both the qubit and the high-$Q$ cavity perform well, with qubit $T_{1}=110$\,$\upmu$s, qubit Ramsey decay time $T_{2}^{*}=40$\,$\upmu$s, cavity $T_{1}=2.8$\,ms (Figure~\ref{integration}b), and cavity $T_{2}^{*}=1.5$\,ms for the $\ket{1}$ Fock state\cite{reagor_quantum_2015}. These qubit lifetimes are among the best measured in 3D cavities, and the coaxial stub resonator $T_{1}$ does not decrease when a qubit is added. This suggests that no additional sources of dissipation are introduced when these elements are combined into a single, seamless package.

Integration with 3D cavities is not strictly necessary to produce a module with many coherent circuit elements. In an all-lithographic system on a single chip, we can add components without sacrificing the isolation between non-adjacent elements. Figure~\ref{integration}c shows a coax-line variant with two additional stripline resonators. These resonators function as a bandpass Purcell filter\cite{reed_fast_2010,jeffrey_fast_2014} and high-$Q$ storage resonator. We measure coherence consistent with qubit-and-stripline designs (best qubit $T_{1}\approx 60$\,$\upmu$s, $T_{2}^{*}\approx 50$\,$\upmu$s, Hahn echo decay time $T_{2}\approx 60$\,$\upmu$s) and with a large cavity decay rate and qubit-readout coupling. Numerous devices were measured, producing consistently long lifetimes (Table~\ref{seammux}). The best stripline storage resonator in this four-element module has $T_{1} = 250$\,$\upmu$s, or an equivalent $\Q{i}=11.2$ million. These results demonstrate that entirely chip-based designs can produce a highly coherent quantum module. A concept for how these modules could be expanded, including eight Purcell-filtered qubits and two multiplexed readout lines, is shown in Figure~\ref{integration}d. 

\section{Outlook and Conclusion}

We have introduced a 3D enclosure and chip-based coax-line architecture that allow for complex \ac{cQED} experiments. The device construction, absent of seams and based on natural waveguide isolation, suppresses spurious modes and allows for precisely-engineered couplings. Using undercoupled stripline resonators, we measured the highest $\Q{i}$ in a chip-based \ac{cQED} device to date. We integrated this type of resonator with a qubit and a millisecond 3D cavity without introducing further loss, creating a long-lived, multi-cavity \ac{cQED} system. Adaptations of this system have already been used in more complex experiments with multiple qubits or cavities\cite{chou_implementing_????,wang_schrodinger_????}.

Using this architecture, existing designs may be improved and new capabilities created. On-chip amplification schemes could possibly be integrated with readout feedlines, allowing multiplexed single-shot readout. Many more elements could be added, producing dense multi-qubit, multi-cavity systems. The concept could apply to wafer-scale micromachining designs, where more complex, multi-layer circuits could be fabricated\cite{brecht_multilayer_2016}.

This flexibility, in addition to high coherence properties, may yet inspire the next generation of hardware towards fault-tolerant error correction.

\section*{Acknowledgments}

We thank M. H. Devoret, J. Blumoff, K. Chou and E. Holland for helpful discussions and T. Brecht for technical assistance. This research was supported by the U.S. Army Research Office (W911NF-14-1-0011). Facilities use was supported by the Yale Institute for Nanoscience and Quantum Engineering (YINQE), the Yale SEAS cleanroom, and the NSF (MRSECDMR 1119826). C.A. acknowledges support from the NSF Graduate Research Fellowship under Grant No. DGE-1122492. K. S. acknowledges support from the Yale Science Scholars Fellowship. W. P. was supported by NSF grant PHY1309996 and by a fellowship instituted with a Max Planck Research Award from the Alexander von Humboldt Foundation.


\renewcommand{\thesection}{S\arabic{section}}
\renewcommand{\thefigure}{S\arabic{figure}}
\renewcommand{\thetable}{S\arabic{table}}
\setcounter{section}{0}
\setcounter{figure}{0}
\setcounter{table}{0}

\section*{Supplementary Material}
\section{Fabrication and Assembly}
Resonators were fabricated on c-plane EFG sapphire chips using photolithography, deposition of evaporated aluminum 80\,nm thick, and a lift-off process. Resonators with qubits were fabricated simultaneously using electron beam lithography, bridge-free double-angle evaporation\cite{lecocq_junction_2011}, and a lift-off process. No significant difference in $\Q{i}$ was noted between resonators fabricated using the two methods. The enclosure was machined from high-purity (99.995\%) aluminum and chemically etched using Aluminum Etchant A (Transene, Danvers, MA) for 4 hours. Alignment within the waveguide was achieved mechanically. Coupling pin lengths were chosen in a manner identical to that of standard 3D cavities.

\section{Measurement}
Qubits were measured in transmission using asymmetric coupling (a weakly-coupled input and a strongly-coupled output port) and standard high-power readout techniques\cite{reed_entanglement_2013}. Resonators were measured in transmission or feedline-coupled configurations. In the feedline-coupled configuration, the measurement transmission line is interrupted by a tee connected to one port of the resonator. Differing impedance in each direction as seen from the device port introduces an asymmetry in resonance lineshape. (This configuration is sometimes called ``hanger'' since the resonator appears to hang off the feedline.) Resonator quality factors were extracted by fitting to a complex model for $S_{21}$\cite{khalil_analysis_2012}, accounting for impedance asymmetry, amplification and attenuation, and electrical delay. Unlike the transmission configuration, feedline-coupled measurements produce internal quality factor $\Q{i}$ and coupling quality factor $\Q{c}$ without knowledge of line losses. Fits were performed in two steps, as illustrated in Figure~2. The standard error for parameters in each fit was propagated to calculate the error in $\Q{i}$ and $\Q{c}$. Measurements at higher powers helped to establish a trend in $\Q{i}$ versus power with which low-power measurements were consistent. To further confirm the measured values of total quality factor, cavity ringdowns were measured following a pulsed excitation, and in the case of resonators with qubits, a cavity $T_{1}$ experiment was performed. Measurements used standard microwave lines, including a HEMT amplifier, in a dilution refrigerator with a measurement stage temperature of about 20\,mK.

\section{Cavity-Qubit System Parameters}
Table~\ref{table} includes notable device parameters. The $\kappa_{\text{q,r}}$ associated with the qubit and resonator, respectively, reflect the dissipation-free case, and can be considered the qubit's Purcell limit and resonator's coupling $\Q{c}$. Any addition of loss will increase these values, leading to a large ``deviation''.

Simulation was generally able to predict first-order parameters within several percent of measured values. Larger deviation can be attributed to oxidation of the Josephson junction between measurement of its normal state room temperature resistance and cold measurement. This produces a discrepancy between measured and simulated qubit frequency, which propagates to other parameters. Other errors can be attributed to variability in machining, etching, and mechanical positioning. Improvement in mounting technique should improve parameter accuracy. Finite-element simulations were performed using Ansys HFSS.

\begin{table}[tbp]
\caption{Simulated and measured parameters for the resonator-qubit system corresponding with Figure~3a. The deviation of several parameters is larger than expected due to junction aging (see text). Simulation does not account for dissipation, and so the simulated $\kappa_{\text{r}}$ represents the qubit's Purcell limit.}
\centering  
\begin{tabular}{c c c c} 
\hline\hline                        
Parameter	& Experiment (MHz) 		& Simulation (MHz)		& Deviation (\%) \\
\hline
$\omega_{\text{q}}/2\pi$	& 5441.9				& 5828.3				& 6.6 \\
$\omega_{\text{r}}/2\pi$	& 9269.5				& 9338.0				& 0.7 \\
$\chi_{\text{qr}}/2\pi$		& -2.31					& -2.74					& 15.7 \\
$\chi_{\text{qq}}/2\pi$  	& -238.5				& -236.9				& 0.7 \\
$\chi_{\text{rr}}/2\pi$  	& -						& $-8\times 10^{-3}$ 	& -	\\
$\kappa_{\text{r}}/2\pi$  	& $3.7\times 10^{-3}$ 	& - 					& - \\
$\kappa_{\text{q}}/2\pi$  	& $1.4\times 10^{-2}$ 	& $3.6\times 10^{-5}$ 	& - \\
$p_{\text{e}}$ 				& 2\% 					& - 					& - \\
\hline
Parameter  					& Time ($\upmu$s)		& 						& 	\\
\hline
$T_{1}$  					& 69.9					& 						& 	\\
$T_{2}^{*}$					& 21.4					& 						& 	\\ 
$T_{2}$						& 29.2					& 						&	\\ 
\hline\hline 
\end{tabular}
\label{table}
\end{table}

Table~\ref{seammux} lists important measured parameters for devices of the type shown in Figure~\ref{integration}c. Multiple non-interacting devices were enclosed in the same package, sharing a feedline for multiplexed readout. Therefore, statistics for a relatively large number of devices were obtained. Lower readout $\Q{c}$ (higher $\kappa_{\text{r}}$) seems to correspond with lower values for storage cavity $T_{1}$, consistent with possible Purcell limitation.

\begin{table*}[tbp]
\caption{Measured parameters for separate devices containing readout cavity, qubit, Purcell filter and storage cavity as shown in Figure~\ref{integration}c. Qubit, storage, and readout resonant frequencies ($\omega_{\text{q,s,r}}$) as well as parameters $\chi$ and $\kappa$ are divided by $2\pi$ and shown in units of MHz. Qubit coherence times $T_{1}$, $T_{2}^{*}$, and $T_{2}$ are in $\upmu$s, and are the average of measurements over several hours with a typical spread of $\pm$10\,$\upmu$s. The highest individual $T_{1}=85$\,$\upmu$s, $T_{2}^{*}=67$\,$\upmu$s, and $T_{2}=80$\,$\upmu$s. Omitted values were not measured.}
\centering
\begin{tabular}{c|c|c|c|c|c|c|c|c|c|c|c} 
\hline\hline
Device & $\omega_{\text{q}}$ (MHz) & $\omega_{\text{s}}$ (MHz) & $\omega_{\text{r}}$ (MHz) & $\chi_{\text{qr}}$ (MHz) & $\chi_{\text{qs}}$ (MHz) & $\kappa_{\text{r}}$ (MHz) & $T_{1}$ ($\upmu$s) & $T_{2}^{*}$ ($\upmu$s) & $T_{2}$ ($\upmu$s) & $T_{1}^{\text{s}}$ ($\upmu$s) & $\Q{i}^{\text{s}}$ ($\times 10^{6}$)\\
\hline
1A 	& 5509.3 	& 6990.0 	& 9350.9 	& 3.10 	& 1.84 	& 4.14 	& 47	& 28	& 62	& 37	& 1.63\\
1C 	& 5431.1 	& 7125.7 	& 9406.5 	& - 	& 2.00 	& 0.34 	& 68	& 26	& 41	& 143	& 6.40\\
2B 	& 5423.7 	& 7035.1 	& 9377.7 	& 2.55 	& 1.76 	& 4.50 	& 43 	& 18 	& 47	& 37	& 1.64\\
2C 	& 5362.8 	& 7124.8 	& 9436.1 	& - 	& 1.34 	& 0.46 	& 51 	& 13 	& 78	& 85	& 3.81\\
3A 	& 5573.9 	& 7065.5 	& 9447.5 	& 2.38 	& 1.85 	& 2.88 	& 62 	& 9 	& 51	& 121	& 5.37\\
3B 	& 5345.8 	& 7066.2 	& 9442.8 	& 2.47 	& 1.59 	& 2.15	& 62 	& 49 	& 59	& 91	& 4.04\\
3C 	& 5378.8 	& 7160.7 	& 9493.3 	& - 	& 1.58 	& 1.02	& 72 	& 23 	& 69	& 250	& 11.25\\
4A 	& 5551.2 	& 7093.6 	& 9464.7 	& 2.64 	& 1.97 	& 5.43 	& 55 	& 33	& 41	& 112	& 4.99\\
4B 	& 5417.2 	& 7098.2 	& 9475.0 	& 1.60 	& 1.59 	& 5.89 	& 41 	& 15	& 56	& 121	& 5.40\\
4C 	& 5269.8 	& 7158.0 	& 9511.7 	& - 	& 1.35 	& 0.27	& 13 	& 0.2	& 4		& 154	& 6.93\\
5A 	& 5672.8 	& 7108.0 	& 9542.1 	& 1.62 	& 1.97 	& 9.47 	& 23 	& 21 	& 20	& 25	& 1.12\\
5B 	& 5540.8 	& 7098.0 	& 9536.0 	& 0.08 	& 1.68 	& 8.65 	& 26 	& 18	& 21	& 66	& 2.94\\
5C 	& 5459.6 	& 7228.0 	& 9607.2 	& - 	& 1.67 	& 2.57 	& 36 	& 14	& 33	& 154	& 6.99\\
\hline\hline
\end{tabular}
\label{seammux}
\end{table*}

\section{Loss Determination}
To determine whether the resonator loss can be explained using conventional theories, we measure the quality factors (and qubit lifetimes, when present) of nearly 100 devices and performed corresponding loss simulations. We consider several likely dissipation sources, including bulk substrate dielectric, interface dielectrics, and conductor surface effects. The electric fields of both resonator and qubit modes fill similar fractions of dielectric as their analogues in standard 3D cavities, so we expect them to be limited by similar mechanisms. We refer to the fraction of the electric field energy stored in the volume labeled $\text{i}$ as the participation ratio $\particip{i}$\cite{wang_surface_2015}.

\begin{figure}[tbp]
  \centering
  \includegraphics[width=1.0\linewidth]{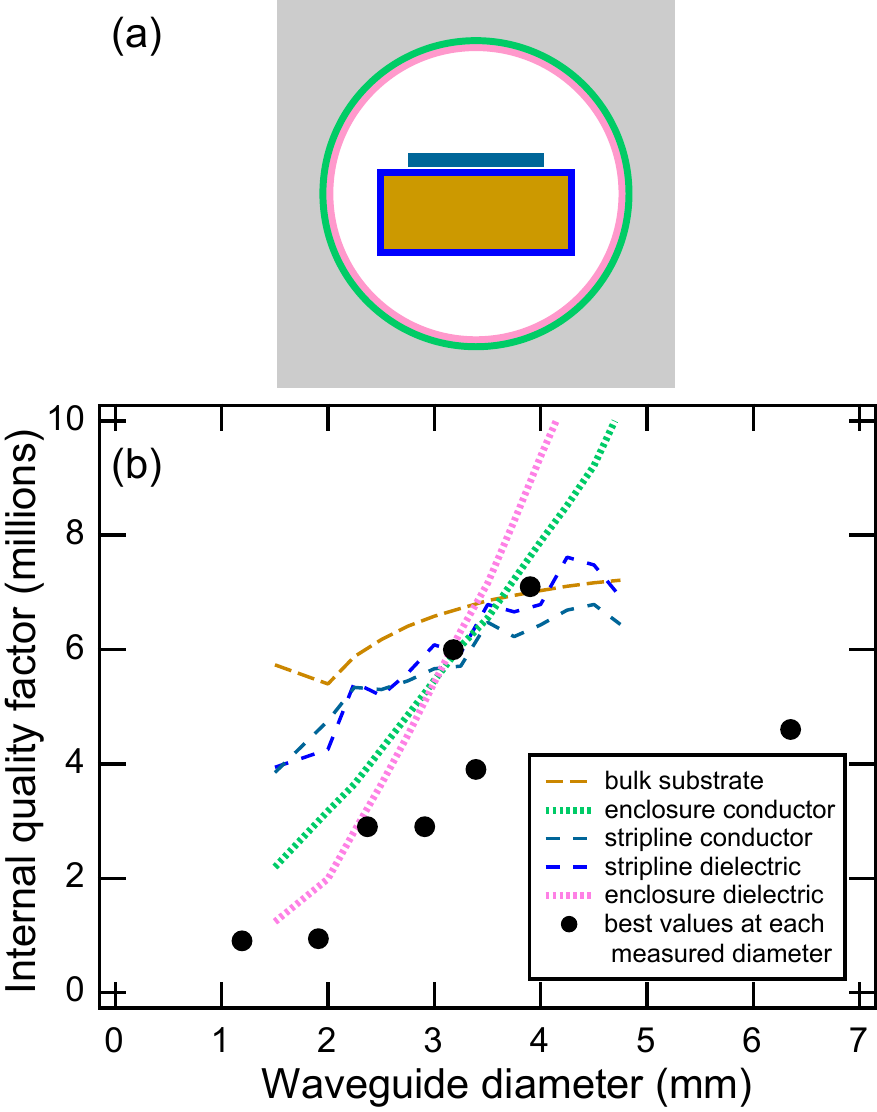}
  \caption{(a) A schematic cross-section of the enclosure, substrate, and stripline. Sizes are exaggerated. The colored lines suggest possible dissipation sources (see text) and correspond with the legend in (b). (b) Stripline resonators with similar designs were measured in packages with different waveguide diameters. At most diameters, more than one sample was measured. The highest value at each diameter is plotted (black circles). Participation ratios $\particip{i}$ are simulated for several loss sources. The corresponding limiting quality factors, obtained by $\Q{limit}=\Q{material}/\particip{i}$ for each material, are calculated and plotted using dashed or dotted lines. Each $\Q{limit}$ trace is scaled by a value of $\Q{material}$ to lie at or above the highest data point, representing a conservative bound. The dashed lines, relating to $\particip{i}$ of bulk substrate (orange), stripline conductor surface (turquoise), and stripline dielectric layers (blue), are not very consistent with the measured behavior. The dotted lines, representing the enclosure conductor surface (green) and any enclosure dielectric (pink), scale more appropriately. However, the values of $\Q{material}$ used to scale these traces are considerably lower than the bounding values of $\Q{material}$ determined empirically in other experiments. Noisy and jagged features in the simulated data are numerical anomalies due to difficulty of combining large 3D simulations with small planar features.}
     \label{vsdiam}
\end{figure}

We obtain the largest dynamic range in participation by varying the diameter of waveguide enclosures with diameters ranging from about 1--6\,mm. Internal quality factor are observed to increase from about 1 to 8 million as the waveguide diameter increased, and then taper off, with the largest values measured for a diameter equal to 4\,mm (Figure~\ref{vsdiam}). At very large diameters, reduced waveguide isolation may introduce radiative loss or increase the field's sampling of seams at the waveguide ends. A larger mode volume is consistent with reduced electric field participation at surfaces. Unfortunately, as in former studies\cite{wang_surface_2015,dial_bulk_2015,sandberg_etch_2012}, loss among the contributors scales similarly between each surface as a function of enclosure diameter.

The best resonator internal quality factor, $8.0 \pm 0.5$ million, had a larger-than-nominal stripline width and is not included in Figure~\ref{vsdiam}. For similar reasons, resonators measured on chips with qubits, including the stripline storage resonator with $\Q{i}=11.3 \pm 0.3$ million, are not compared in this figure.

From the behavior of five possible contributors---the surface dielectric and conductor surface loss of (1) the enclosure aluminum and (2) the deposited aluminum and chip surfaces, plus the sapphire substrate dielectric---we determine the ones that best fit the trend and analyze them more closely. The strongest diameter-dependent scaling behavior belongs to the waveguide enclosure surface losses. We can use our participation ratios $\particip{i}$ to develop lower bounds on $\Q{material}$ and compare those to established values. Conversely, we can apply the established values of $\Q{material}$ to our $p_i$ to obtain a lower bound on $\Q{i}$. These values are calculated in Table~\ref{loss}, and uses our highest measured $\Q{i}$'s to place the most stringent bounds.

\begin{table*}[tbp]
\caption{Simulated and measured values for system dissipation sources. Values of $\Q{material}$ are lower bounds and are obtained from single-photon measurements of similar systems. Our range of participations $\particip{i}$ are obtained from simulation, except for conductor participation (where $\particip{i} = \alpha$, kinetic inductance fraction), which is measured. The calculation of our $\Q{material}$ comes from the best measured $\Q{i}=8$ million, while the calculated $\Q{i}$ limit is a product of the established $\Q{material}$ bound and $1/\particip{i}$.}
\centering
\begin{tabular}{c c c c c c} 
\hline\hline                        
Source & established $\Q{material}$ & $\Q{material}$ & our $\particip{i}$ & our $\Q{material}$ & our $\Q{i}$ limit \\
 & bound & source &  & bound & $\times 10^6$ \\
\hline
Substrate sapphire & 1--5 $\times 10^6$ & 3D qubits\cite{paik_observation_2011,rigetti_superconducting_2012}, WGMR\cite{creedon_high_2011} & 0.4--0.5 & 3.2 $\times 10^6$ & 2.0--12.5 \\
Enclosure conductor & $4400$ & cylindrical cavity\cite{reagor_reaching_2013} & 1--10 $\times 10^{-6}$ & 8 & 440--4400 \\
Stripline conductor & $4800$ & Al WGMR\cite{minev_planar_2013} & 1--4 $\times 10^{-3}$ & 8000 & 1.2--4.8 \\
Stripline dielectric interfaces & $380$ & 3D qubits\cite{wang_surface_2015} & 1--3 $\times 10^{-5}$ & 80 & 13--38 \\
Enclosure dielectric & $750$ & coaxial stub cavity\cite{reagor_quantum_2015} & 8 $\times 10^{-7}$ & 6.4 & 940 \\
\hline\hline
\end{tabular}
\label{loss}
\end{table*}

The best bound on the dielectric loss tangent of similar-quality sapphire comes from 3D qubits with lifetimes exceeding 100\,$\upmu$s. These qubit modes store 90\% of their electric field energy in the sapphire substrate, and appear to be limited by other loss mechanisms\cite{wang_surface_2015}. Experiments in separate systems with higher quality sapphire have determined even higher bounds\cite{creedon_high_2011}. Table~\ref{loss} shows the values obtained for our standard devices. However, by using a non-standard device, we can alter the $\particip{i}$ for sapphire substrate participation and place a higher bound. In one instance, we placed two stripline resonators side-by-side on the same substrate. In another, the two resonators were patterned on the front and back of the substrate. These variations allowed a differential and a common mode with significantly different $\particip{i}$, ranging from 24--89\%. For each device, the mode with higher sapphire participation exhibited a $\Q{i}$ about 50\% higher than the opposite mode. This inverse relationship does not identify the dominant loss source, but does seem to rule out our substrate material.

Next, we considered the kinetic inductance fraction $\alpha$---a metric of dissipation within the superconductor---and measured its effect using the dependence of resonant frequency on temperature\cite{reagor_reaching_2013}. Repeating this experiment with striplines made from higher-$T_{\text{c}}$ niobium, we obtained the $\alpha$ of the waveguide enclosure independently. Our standard design resulted in $\alpha=1\mbox{--}4\times 10^{-3}$ for the stripline and $\alpha=10^{-6}\mbox{--}10^{-5}$ for the waveguide enclosure. Using the established $\Q{material}$ bound on etched high-purity bulk aluminum\cite{reagor_reaching_2013} we place a conservative lower limit on $\Q{i}$ due to waveguide surface dissipation at $4 \times 10^{8}$. Even though the diameter-dependent behavior of waveguide conductor surface loss in Figure~\ref{vsdiam} matches well with measured $\Q{i}$, it seems highly unlikely that this is the limiting loss source based on formerly established bounds.

We consider thin-film deposited aluminum separately, since it undergoes fundamentally different processing. Here, the best published bounds\cite{minev_planar_2013} predict this loss mechanism to be strongly limiting. However, an updated study has established significantly higher bounds\cite{serniak_superconducting_????}. Additionally, the value of $\alpha$ is not measured to change significantly as waveguide diameter is varied, and so cannot explain the measured changes in $\Q{i}$ with diameter.

Loss contribution from surface dielectric interfaces (at the metal-substrate, substrate-vacuum, or metal-vacuum interfaces) remain a plausible limit. The best established limit on those surfaces\cite{wang_surface_2015} is within an order of magnitude of our value. However, as with the other surface effects, the diameter-dependent scaling is not consistent with measured $\Q{i}$. This comparison is further obscured by the fact that the relative contribution of these three primary surfaces cannot be empirically differentiated.

While no single source conclusively dominates the loss in our system, we may be limited my many sources contributing at similar levels. Alternatively, we may be affected by loss from an unconsidered source, such as vortices, mechanical dampening, or substrate processing. The strong dependence of $\Q{i}$ on waveguide diameter is suggestive of a non-localized material dissipation, and is consistent with prior observations of $\Q{i}$ proportional to mode volume in cQED systems. The ability of this system to controllably evaluate on-chip and package losses independently, to some degree, lends versatility to the design as a testbed for losses. Future studies may be able to place and raise these bounds more definitively.

\section*{References}
%


\begin{thebibliography}{33}%
\makeatletter
\providecommand \@ifxundefined [1]{%
 \@ifx{#1\undefined}
}%
\providecommand \@ifnum [1]{%
 \ifnum #1\expandafter \@firstoftwo
 \else \expandafter \@secondoftwo
 \fi
}%
\providecommand \@ifx [1]{%
 \ifx #1\expandafter \@firstoftwo
 \else \expandafter \@secondoftwo
 \fi
}%
\providecommand \natexlab [1]{#1}%
\providecommand \enquote  [1]{``#1''}%
\providecommand \bibnamefont  [1]{#1}%
\providecommand \bibfnamefont [1]{#1}%
\providecommand \citenamefont [1]{#1}%
\providecommand \href@noop [0]{\@secondoftwo}%
\providecommand \href [0]{\begingroup \@sanitize@url \@href}%
\providecommand \@href[1]{\@@startlink{#1}\@@href}%
\providecommand \@@href[1]{\endgroup#1\@@endlink}%
\providecommand \@sanitize@url [0]{\catcode `\\12\catcode `\$12\catcode
  `\&12\catcode `\#12\catcode `\^12\catcode `\_12\catcode `\%12\relax}%
\providecommand \@@startlink[1]{}%
\providecommand \@@endlink[0]{}%
\providecommand \url  [0]{\begingroup\@sanitize@url \@url }%
\providecommand \@url [1]{\endgroup\@href {#1}{\urlprefix }}%
\providecommand \urlprefix  [0]{URL }%
\providecommand \Eprint [0]{\href }%
\providecommand \doibase [0]{http://dx.doi.org/}%
\providecommand \selectlanguage [0]{\@gobble}%
\providecommand \bibinfo  [0]{\@secondoftwo}%
\providecommand \bibfield  [0]{\@secondoftwo}%
\providecommand \translation [1]{[#1]}%
\providecommand \BibitemOpen [0]{}%
\providecommand \bibitemStop [0]{}%
\providecommand \bibitemNoStop [0]{.\EOS\space}%
\providecommand \EOS [0]{\spacefactor3000\relax}%
\providecommand \BibitemShut  [1]{\csname bibitem#1\endcsname}%
\let\auto@bib@innerbib\@empty
\bibitem [{\citenamefont {Barends}\ \emph {et~al.}(2014)\citenamefont
  {Barends}, \citenamefont {Kelly}, \citenamefont {Megrant}, \citenamefont
  {Veitia}, \citenamefont {Sank}, \citenamefont {Jeffrey}, \citenamefont
  {White}, \citenamefont {Mutus}, \citenamefont {Fowler}, \citenamefont
  {Campbell}, \citenamefont {Chen}, \citenamefont {Chen}, \citenamefont
  {Chiaro}, \citenamefont {Dunsworth}, \citenamefont {Neill}, \citenamefont
  {O’Malley}, \citenamefont {Roushan}, \citenamefont {Vainsencher},
  \citenamefont {Wenner}, \citenamefont {Korotkov}, \citenamefont {Cleland},\
  and\ \citenamefont {Martinis}}]{barends_superconducting_2014}%
  \BibitemOpen
  \bibfield  {author} {\bibinfo {author} {\bibfnamefont {R.}~\bibnamefont
  {Barends}}, \bibinfo {author} {\bibfnamefont {J.}~\bibnamefont {Kelly}},
  \bibinfo {author} {\bibfnamefont {A.}~\bibnamefont {Megrant}}, \bibinfo
  {author} {\bibfnamefont {A.}~\bibnamefont {Veitia}}, \bibinfo {author}
  {\bibfnamefont {D.}~\bibnamefont {Sank}}, \bibinfo {author} {\bibfnamefont
  {E.}~\bibnamefont {Jeffrey}}, \bibinfo {author} {\bibfnamefont {T.~C.}\
  \bibnamefont {White}}, \bibinfo {author} {\bibfnamefont {J.}~\bibnamefont
  {Mutus}}, \bibinfo {author} {\bibfnamefont {A.~G.}\ \bibnamefont {Fowler}},
  \bibinfo {author} {\bibfnamefont {B.}~\bibnamefont {Campbell}}, \bibinfo
  {author} {\bibfnamefont {Y.}~\bibnamefont {Chen}}, \bibinfo {author}
  {\bibfnamefont {Z.}~\bibnamefont {Chen}}, \bibinfo {author} {\bibfnamefont
  {B.}~\bibnamefont {Chiaro}}, \bibinfo {author} {\bibfnamefont
  {A.}~\bibnamefont {Dunsworth}}, \bibinfo {author} {\bibfnamefont
  {C.}~\bibnamefont {Neill}}, \bibinfo {author} {\bibfnamefont
  {P.}~\bibnamefont {O’Malley}}, \bibinfo {author} {\bibfnamefont
  {P.}~\bibnamefont {Roushan}}, \bibinfo {author} {\bibfnamefont
  {A.}~\bibnamefont {Vainsencher}}, \bibinfo {author} {\bibfnamefont
  {J.}~\bibnamefont {Wenner}}, \bibinfo {author} {\bibfnamefont {A.~N.}\
  \bibnamefont {Korotkov}}, \bibinfo {author} {\bibfnamefont {A.~N.}\
  \bibnamefont {Cleland}}, \ and\ \bibinfo {author} {\bibfnamefont {J.~M.}\
  \bibnamefont {Martinis}},\ }\href {\doibase 10.1038/nature13171} {\bibfield
  {journal} {\bibinfo  {journal} {Nature}\ }\textbf {\bibinfo {volume} {508}},\
  \bibinfo {pages} {500} (\bibinfo {year} {2014})}\BibitemShut {NoStop}%
\bibitem [{\citenamefont {Steane}(2003)}]{steane_overhead_2003}%
  \BibitemOpen
  \bibfield  {author} {\bibinfo {author} {\bibfnamefont {A.~M.}\ \bibnamefont
  {Steane}},\ }\href {\doibase 10.1103/PhysRevA.68.042322} {\bibfield
  {journal} {\bibinfo  {journal} {Physical Review A}\ }\textbf {\bibinfo
  {volume} {68}},\ \bibinfo {pages} {042322} (\bibinfo {year}
  {2003})}\BibitemShut {NoStop}%
\bibitem [{\citenamefont {Gambetta}, \citenamefont {Chow},\ and\ \citenamefont
  {Steffen}(2015)}]{gambetta_building_2015}%
  \BibitemOpen
  \bibfield  {author} {\bibinfo {author} {\bibfnamefont {J.~M.}\ \bibnamefont
  {Gambetta}}, \bibinfo {author} {\bibfnamefont {J.~M.}\ \bibnamefont {Chow}},
  \ and\ \bibinfo {author} {\bibfnamefont {M.}~\bibnamefont {Steffen}},\ }\href
  {http://arxiv.org/abs/1510.04375} {\bibfield  {journal} {\bibinfo  {journal}
  {arXiv:1510.04375 [quant-ph]}\ } (\bibinfo {year} {2015})}\BibitemShut
  {NoStop}%
\bibitem [{\citenamefont {Brecht}\ \emph {et~al.}(2016)\citenamefont {Brecht},
  \citenamefont {Pfaff}, \citenamefont {Wang}, \citenamefont {Chu},
  \citenamefont {Frunzio}, \citenamefont {Devoret},\ and\ \citenamefont
  {Schoelkopf}}]{brecht_multilayer_2016}%
  \BibitemOpen
  \bibfield  {author} {\bibinfo {author} {\bibfnamefont {T.}~\bibnamefont
  {Brecht}}, \bibinfo {author} {\bibfnamefont {W.}~\bibnamefont {Pfaff}},
  \bibinfo {author} {\bibfnamefont {C.}~\bibnamefont {Wang}}, \bibinfo {author}
  {\bibfnamefont {Y.}~\bibnamefont {Chu}}, \bibinfo {author} {\bibfnamefont
  {L.}~\bibnamefont {Frunzio}}, \bibinfo {author} {\bibfnamefont {M.~H.}\
  \bibnamefont {Devoret}}, \ and\ \bibinfo {author} {\bibfnamefont {R.~J.}\
  \bibnamefont {Schoelkopf}},\ }\href {\doibase 10.1038/npjqi.2016.2}
  {\bibfield  {journal} {\bibinfo  {journal} {npj Quantum Information}\
  }\textbf {\bibinfo {volume} {2}},\ \bibinfo {pages} {16002} (\bibinfo {year}
  {2016})}\BibitemShut {NoStop}%
\bibitem [{\citenamefont {Chen}\ \emph {et~al.}(2014)\citenamefont {Chen},
  \citenamefont {Megrant}, \citenamefont {Kelly}, \citenamefont {Barends},
  \citenamefont {Bochmann}, \citenamefont {Chen}, \citenamefont {Chiaro},
  \citenamefont {Dunsworth}, \citenamefont {Jeffrey}, \citenamefont {Mutus},
  \citenamefont {O'Malley}, \citenamefont {Neill}, \citenamefont {Roushan},
  \citenamefont {Sank}, \citenamefont {Vainsencher}, \citenamefont {Wenner},
  \citenamefont {White}, \citenamefont {Cleland},\ and\ \citenamefont
  {Martinis}}]{chen_fabrication_2014}%
  \BibitemOpen
  \bibfield  {author} {\bibinfo {author} {\bibfnamefont {Z.}~\bibnamefont
  {Chen}}, \bibinfo {author} {\bibfnamefont {A.}~\bibnamefont {Megrant}},
  \bibinfo {author} {\bibfnamefont {J.}~\bibnamefont {Kelly}}, \bibinfo
  {author} {\bibfnamefont {R.}~\bibnamefont {Barends}}, \bibinfo {author}
  {\bibfnamefont {J.}~\bibnamefont {Bochmann}}, \bibinfo {author}
  {\bibfnamefont {Y.}~\bibnamefont {Chen}}, \bibinfo {author} {\bibfnamefont
  {B.}~\bibnamefont {Chiaro}}, \bibinfo {author} {\bibfnamefont
  {A.}~\bibnamefont {Dunsworth}}, \bibinfo {author} {\bibfnamefont
  {E.}~\bibnamefont {Jeffrey}}, \bibinfo {author} {\bibfnamefont {J.~Y.}\
  \bibnamefont {Mutus}}, \bibinfo {author} {\bibfnamefont {P.~J.~J.}\
  \bibnamefont {O'Malley}}, \bibinfo {author} {\bibfnamefont {C.}~\bibnamefont
  {Neill}}, \bibinfo {author} {\bibfnamefont {P.}~\bibnamefont {Roushan}},
  \bibinfo {author} {\bibfnamefont {D.}~\bibnamefont {Sank}}, \bibinfo {author}
  {\bibfnamefont {A.}~\bibnamefont {Vainsencher}}, \bibinfo {author}
  {\bibfnamefont {J.}~\bibnamefont {Wenner}}, \bibinfo {author} {\bibfnamefont
  {T.~C.}\ \bibnamefont {White}}, \bibinfo {author} {\bibfnamefont {A.~N.}\
  \bibnamefont {Cleland}}, \ and\ \bibinfo {author} {\bibfnamefont {J.~M.}\
  \bibnamefont {Martinis}},\ }\href {\doibase 10.1063/1.4863745} {\bibfield
  {journal} {\bibinfo  {journal} {Applied Physics Letters}\ }\textbf {\bibinfo
  {volume} {104}},\ \bibinfo {pages} {052602} (\bibinfo {year}
  {2014})}\BibitemShut {NoStop}%
\bibitem [{\citenamefont {Brecht}\ \emph {et~al.}(2015)\citenamefont {Brecht},
  \citenamefont {Reagor}, \citenamefont {Chu}, \citenamefont {Pfaff},
  \citenamefont {Wang}, \citenamefont {Frunzio}, \citenamefont {Devoret},\ and\
  \citenamefont {Schoelkopf}}]{brecht_demonstration_2015}%
  \BibitemOpen
  \bibfield  {author} {\bibinfo {author} {\bibfnamefont {T.}~\bibnamefont
  {Brecht}}, \bibinfo {author} {\bibfnamefont {M.}~\bibnamefont {Reagor}},
  \bibinfo {author} {\bibfnamefont {Y.}~\bibnamefont {Chu}}, \bibinfo {author}
  {\bibfnamefont {W.}~\bibnamefont {Pfaff}}, \bibinfo {author} {\bibfnamefont
  {C.}~\bibnamefont {Wang}}, \bibinfo {author} {\bibfnamefont {L.}~\bibnamefont
  {Frunzio}}, \bibinfo {author} {\bibfnamefont {M.~H.}\ \bibnamefont
  {Devoret}}, \ and\ \bibinfo {author} {\bibfnamefont {R.~J.}\ \bibnamefont
  {Schoelkopf}},\ }\href {\doibase 10.1063/1.4935541} {\bibfield  {journal}
  {\bibinfo  {journal} {Applied Physics Letters}\ }\textbf {\bibinfo {volume}
  {107}},\ \bibinfo {pages} {192603} (\bibinfo {year} {2015})}\BibitemShut
  {NoStop}%
\bibitem [{\citenamefont {Kirchmair}\ \emph {et~al.}(2013)\citenamefont
  {Kirchmair}, \citenamefont {Vlastakis}, \citenamefont {Leghtas},
  \citenamefont {Nigg}, \citenamefont {Paik}, \citenamefont {Ginossar},
  \citenamefont {Mirrahimi}, \citenamefont {Frunzio}, \citenamefont {Girvin},\
  and\ \citenamefont {Schoelkopf}}]{kirchmair_observation_2013}%
  \BibitemOpen
  \bibfield  {author} {\bibinfo {author} {\bibfnamefont {G.}~\bibnamefont
  {Kirchmair}}, \bibinfo {author} {\bibfnamefont {B.}~\bibnamefont
  {Vlastakis}}, \bibinfo {author} {\bibfnamefont {Z.}~\bibnamefont {Leghtas}},
  \bibinfo {author} {\bibfnamefont {S.~E.}\ \bibnamefont {Nigg}}, \bibinfo
  {author} {\bibfnamefont {H.}~\bibnamefont {Paik}}, \bibinfo {author}
  {\bibfnamefont {E.}~\bibnamefont {Ginossar}}, \bibinfo {author}
  {\bibfnamefont {M.}~\bibnamefont {Mirrahimi}}, \bibinfo {author}
  {\bibfnamefont {L.}~\bibnamefont {Frunzio}}, \bibinfo {author} {\bibfnamefont
  {S.~M.}\ \bibnamefont {Girvin}}, \ and\ \bibinfo {author} {\bibfnamefont
  {R.~J.}\ \bibnamefont {Schoelkopf}},\ }\href {\doibase 10.1038/nature11902}
  {\bibfield  {journal} {\bibinfo  {journal} {Nature}\ }\textbf {\bibinfo
  {volume} {495}},\ \bibinfo {pages} {205} (\bibinfo {year}
  {2013})}\BibitemShut {NoStop}%
\bibitem [{\citenamefont {Devoret}\ and\ \citenamefont
  {Schoelkopf}(2013)}]{devoret_superconducting_2013}%
  \BibitemOpen
  \bibfield  {author} {\bibinfo {author} {\bibfnamefont {M.~H.}\ \bibnamefont
  {Devoret}}\ and\ \bibinfo {author} {\bibfnamefont {R.~J.}\ \bibnamefont
  {Schoelkopf}},\ }\href {\doibase 10.1126/science.1231930} {\bibfield
  {journal} {\bibinfo  {journal} {Science}\ }\textbf {\bibinfo {volume}
  {339}},\ \bibinfo {pages} {1169} (\bibinfo {year} {2013})}\BibitemShut
  {NoStop}%
\bibitem [{\citenamefont {Nickerson}, \citenamefont {Fitzsimons},\ and\
  \citenamefont {Benjamin}(2014)}]{nickerson_freely_2014}%
  \BibitemOpen
  \bibfield  {author} {\bibinfo {author} {\bibfnamefont {N.~H.}\ \bibnamefont
  {Nickerson}}, \bibinfo {author} {\bibfnamefont {J.~F.}\ \bibnamefont
  {Fitzsimons}}, \ and\ \bibinfo {author} {\bibfnamefont {S.~C.}\ \bibnamefont
  {Benjamin}},\ }\href {\doibase 10.1103/PhysRevX.4.041041} {\bibfield
  {journal} {\bibinfo  {journal} {Physical Review X}\ }\textbf {\bibinfo
  {volume} {4}},\ \bibinfo {pages} {041041} (\bibinfo {year}
  {2014})}\BibitemShut {NoStop}%
\bibitem [{\citenamefont {Paik}\ \emph {et~al.}(2011)\citenamefont {Paik},
  \citenamefont {Schuster}, \citenamefont {Bishop}, \citenamefont {Kirchmair},
  \citenamefont {Catelani}, \citenamefont {Sears}, \citenamefont {Johnson},
  \citenamefont {Reagor}, \citenamefont {Frunzio}, \citenamefont {Glazman},
  \citenamefont {Girvin}, \citenamefont {Devoret},\ and\ \citenamefont
  {Schoelkopf}}]{paik_observation_2011}%
  \BibitemOpen
  \bibfield  {author} {\bibinfo {author} {\bibfnamefont {H.}~\bibnamefont
  {Paik}}, \bibinfo {author} {\bibfnamefont {D.~I.}\ \bibnamefont {Schuster}},
  \bibinfo {author} {\bibfnamefont {L.~S.}\ \bibnamefont {Bishop}}, \bibinfo
  {author} {\bibfnamefont {G.}~\bibnamefont {Kirchmair}}, \bibinfo {author}
  {\bibfnamefont {G.}~\bibnamefont {Catelani}}, \bibinfo {author}
  {\bibfnamefont {A.~P.}\ \bibnamefont {Sears}}, \bibinfo {author}
  {\bibfnamefont {B.~R.}\ \bibnamefont {Johnson}}, \bibinfo {author}
  {\bibfnamefont {M.~J.}\ \bibnamefont {Reagor}}, \bibinfo {author}
  {\bibfnamefont {L.}~\bibnamefont {Frunzio}}, \bibinfo {author} {\bibfnamefont
  {L.~I.}\ \bibnamefont {Glazman}}, \bibinfo {author} {\bibfnamefont {S.~M.}\
  \bibnamefont {Girvin}}, \bibinfo {author} {\bibfnamefont {M.~H.}\
  \bibnamefont {Devoret}}, \ and\ \bibinfo {author} {\bibfnamefont {R.~J.}\
  \bibnamefont {Schoelkopf}},\ }\href {\doibase 10.1103/PhysRevLett.107.240501}
  {\bibfield  {journal} {\bibinfo  {journal} {Physical Review Letters}\
  }\textbf {\bibinfo {volume} {107}},\ \bibinfo {pages} {240501} (\bibinfo
  {year} {2011})}\BibitemShut {NoStop}%
\bibitem [{\citenamefont {Dial}\ \emph {et~al.}(2015)\citenamefont {Dial},
  \citenamefont {McClure}, \citenamefont {Poletto}, \citenamefont {Gambetta},
  \citenamefont {Abraham}, \citenamefont {Chow},\ and\ \citenamefont
  {Steffen}}]{dial_bulk_2015}%
  \BibitemOpen
  \bibfield  {author} {\bibinfo {author} {\bibfnamefont {O.}~\bibnamefont
  {Dial}}, \bibinfo {author} {\bibfnamefont {D.~T.}\ \bibnamefont {McClure}},
  \bibinfo {author} {\bibfnamefont {S.}~\bibnamefont {Poletto}}, \bibinfo
  {author} {\bibfnamefont {J.~M.}\ \bibnamefont {Gambetta}}, \bibinfo {author}
  {\bibfnamefont {D.~W.}\ \bibnamefont {Abraham}}, \bibinfo {author}
  {\bibfnamefont {J.~M.}\ \bibnamefont {Chow}}, \ and\ \bibinfo {author}
  {\bibfnamefont {M.}~\bibnamefont {Steffen}},\ }\href
  {http://arxiv.org/abs/1509.03859} {\bibfield  {journal} {\bibinfo  {journal}
  {arXiv:1509.03859 [cond-mat, physics:quant-ph]}\ } (\bibinfo {year}
  {2015})}\BibitemShut {NoStop}%
\bibitem [{\citenamefont {Reagor}\ \emph {et~al.}(2015)\citenamefont {Reagor},
  \citenamefont {Pfaff}, \citenamefont {Axline}, \citenamefont {Heeres},
  \citenamefont {Ofek}, \citenamefont {Sliwa}, \citenamefont {Holland},
  \citenamefont {Wang}, \citenamefont {Blumoff}, \citenamefont {Chou},
  \citenamefont {Hatridge}, \citenamefont {Frunzio}, \citenamefont {Devoret},
  \citenamefont {Jiang},\ and\ \citenamefont
  {Schoelkopf}}]{reagor_quantum_2015}%
  \BibitemOpen
  \bibfield  {author} {\bibinfo {author} {\bibfnamefont {M.}~\bibnamefont
  {Reagor}}, \bibinfo {author} {\bibfnamefont {W.}~\bibnamefont {Pfaff}},
  \bibinfo {author} {\bibfnamefont {C.}~\bibnamefont {Axline}}, \bibinfo
  {author} {\bibfnamefont {R.~W.}\ \bibnamefont {Heeres}}, \bibinfo {author}
  {\bibfnamefont {N.}~\bibnamefont {Ofek}}, \bibinfo {author} {\bibfnamefont
  {K.}~\bibnamefont {Sliwa}}, \bibinfo {author} {\bibfnamefont
  {E.}~\bibnamefont {Holland}}, \bibinfo {author} {\bibfnamefont
  {C.}~\bibnamefont {Wang}}, \bibinfo {author} {\bibfnamefont {J.}~\bibnamefont
  {Blumoff}}, \bibinfo {author} {\bibfnamefont {K.}~\bibnamefont {Chou}},
  \bibinfo {author} {\bibfnamefont {M.~J.}\ \bibnamefont {Hatridge}}, \bibinfo
  {author} {\bibfnamefont {L.}~\bibnamefont {Frunzio}}, \bibinfo {author}
  {\bibfnamefont {M.~H.}\ \bibnamefont {Devoret}}, \bibinfo {author}
  {\bibfnamefont {L.}~\bibnamefont {Jiang}}, \ and\ \bibinfo {author}
  {\bibfnamefont {R.~J.}\ \bibnamefont {Schoelkopf}},\ }\href
  {http://arxiv.org/abs/1508.05882} {\bibfield  {journal} {\bibinfo  {journal}
  {arXiv:1508.05882 [cond-mat, physics:quant-ph]}\ } (\bibinfo {year}
  {2015})}\BibitemShut {NoStop}%
\bibitem [{_ap()}]{_apl_????}%
  \BibitemOpen
  \href@noop {} {\ }\bibinfo {note} {See supplemental material for experimental
  details.}\BibitemShut {Stop}%
\bibitem [{\citenamefont {Megrant}\ \emph {et~al.}(2012)\citenamefont
  {Megrant}, \citenamefont {Neill}, \citenamefont {Barends}, \citenamefont
  {Chiaro}, \citenamefont {Chen}, \citenamefont {Feigl}, \citenamefont {Kelly},
  \citenamefont {Lucero}, \citenamefont {Mariantoni}, \citenamefont
  {O’Malley}, \citenamefont {Sank}, \citenamefont {Vainsencher},
  \citenamefont {Wenner}, \citenamefont {White}, \citenamefont {Yin},
  \citenamefont {Zhao}, \citenamefont {Palmstrøm}, \citenamefont {Martinis},\
  and\ \citenamefont {Cleland}}]{megrant_planar_2012}%
  \BibitemOpen
  \bibfield  {author} {\bibinfo {author} {\bibfnamefont {A.}~\bibnamefont
  {Megrant}}, \bibinfo {author} {\bibfnamefont {C.}~\bibnamefont {Neill}},
  \bibinfo {author} {\bibfnamefont {R.}~\bibnamefont {Barends}}, \bibinfo
  {author} {\bibfnamefont {B.}~\bibnamefont {Chiaro}}, \bibinfo {author}
  {\bibfnamefont {Y.}~\bibnamefont {Chen}}, \bibinfo {author} {\bibfnamefont
  {L.}~\bibnamefont {Feigl}}, \bibinfo {author} {\bibfnamefont
  {J.}~\bibnamefont {Kelly}}, \bibinfo {author} {\bibfnamefont
  {E.}~\bibnamefont {Lucero}}, \bibinfo {author} {\bibfnamefont
  {M.}~\bibnamefont {Mariantoni}}, \bibinfo {author} {\bibfnamefont {P.~J.~J.}\
  \bibnamefont {O’Malley}}, \bibinfo {author} {\bibfnamefont
  {D.}~\bibnamefont {Sank}}, \bibinfo {author} {\bibfnamefont {A.}~\bibnamefont
  {Vainsencher}}, \bibinfo {author} {\bibfnamefont {J.}~\bibnamefont {Wenner}},
  \bibinfo {author} {\bibfnamefont {T.~C.}\ \bibnamefont {White}}, \bibinfo
  {author} {\bibfnamefont {Y.}~\bibnamefont {Yin}}, \bibinfo {author}
  {\bibfnamefont {J.}~\bibnamefont {Zhao}}, \bibinfo {author} {\bibfnamefont
  {C.~J.}\ \bibnamefont {Palmstrøm}}, \bibinfo {author} {\bibfnamefont
  {J.~M.}\ \bibnamefont {Martinis}}, \ and\ \bibinfo {author} {\bibfnamefont
  {A.~N.}\ \bibnamefont {Cleland}},\ }\href {\doibase 10.1063/1.3693409}
  {\bibfield  {journal} {\bibinfo  {journal} {Applied Physics Letters}\
  }\textbf {\bibinfo {volume} {100}},\ \bibinfo {pages} {113510} (\bibinfo
  {year} {2012})}\BibitemShut {NoStop}%
\bibitem [{\citenamefont {Sandberg}\ \emph {et~al.}(2013)\citenamefont
  {Sandberg}, \citenamefont {Vissers}, \citenamefont {Ohki}, \citenamefont
  {Gao}, \citenamefont {Aumentado}, \citenamefont {Weides},\ and\ \citenamefont
  {Pappas}}]{sandberg_radiation-suppressed_2013}%
  \BibitemOpen
  \bibfield  {author} {\bibinfo {author} {\bibfnamefont {M.}~\bibnamefont
  {Sandberg}}, \bibinfo {author} {\bibfnamefont {M.~R.}\ \bibnamefont
  {Vissers}}, \bibinfo {author} {\bibfnamefont {T.~A.}\ \bibnamefont {Ohki}},
  \bibinfo {author} {\bibfnamefont {J.}~\bibnamefont {Gao}}, \bibinfo {author}
  {\bibfnamefont {J.}~\bibnamefont {Aumentado}}, \bibinfo {author}
  {\bibfnamefont {M.}~\bibnamefont {Weides}}, \ and\ \bibinfo {author}
  {\bibfnamefont {D.~P.}\ \bibnamefont {Pappas}},\ }\href {\doibase
  10.1063/1.4792698} {\bibfield  {journal} {\bibinfo  {journal} {Applied
  Physics Letters}\ }\textbf {\bibinfo {volume} {102}},\ \bibinfo {pages}
  {072601} (\bibinfo {year} {2013})}\BibitemShut {NoStop}%
\bibitem [{\citenamefont {Khalil}\ \emph {et~al.}(2012)\citenamefont {Khalil},
  \citenamefont {Stoutimore}, \citenamefont {Wellstood},\ and\ \citenamefont
  {Osborn}}]{khalil_analysis_2012}%
  \BibitemOpen
  \bibfield  {author} {\bibinfo {author} {\bibfnamefont {M.~S.}\ \bibnamefont
  {Khalil}}, \bibinfo {author} {\bibfnamefont {M.~J.~A.}\ \bibnamefont
  {Stoutimore}}, \bibinfo {author} {\bibfnamefont {F.~C.}\ \bibnamefont
  {Wellstood}}, \ and\ \bibinfo {author} {\bibfnamefont {K.~D.}\ \bibnamefont
  {Osborn}},\ }\href {\doibase 10.1063/1.3692073} {\bibfield  {journal}
  {\bibinfo  {journal} {Journal of Applied Physics}\ }\textbf {\bibinfo
  {volume} {111}},\ \bibinfo {pages} {054510} (\bibinfo {year}
  {2012})}\BibitemShut {NoStop}%
\bibitem [{\citenamefont {Gao}\ \emph {et~al.}(2008)\citenamefont {Gao},
  \citenamefont {Daal}, \citenamefont {Vayonakis}, \citenamefont {Kumar},
  \citenamefont {Zmuidzinas}, \citenamefont {Sadoulet}, \citenamefont {Mazin},
  \citenamefont {Day},\ and\ \citenamefont {Leduc}}]{gao_experimental_2008}%
  \BibitemOpen
  \bibfield  {author} {\bibinfo {author} {\bibfnamefont {J.}~\bibnamefont
  {Gao}}, \bibinfo {author} {\bibfnamefont {M.}~\bibnamefont {Daal}}, \bibinfo
  {author} {\bibfnamefont {A.}~\bibnamefont {Vayonakis}}, \bibinfo {author}
  {\bibfnamefont {S.}~\bibnamefont {Kumar}}, \bibinfo {author} {\bibfnamefont
  {J.}~\bibnamefont {Zmuidzinas}}, \bibinfo {author} {\bibfnamefont
  {B.}~\bibnamefont {Sadoulet}}, \bibinfo {author} {\bibfnamefont {B.~A.}\
  \bibnamefont {Mazin}}, \bibinfo {author} {\bibfnamefont {P.~K.}\ \bibnamefont
  {Day}}, \ and\ \bibinfo {author} {\bibfnamefont {H.~G.}\ \bibnamefont
  {Leduc}},\ }\href {\doibase 10.1063/1.2906373} {\bibfield  {journal}
  {\bibinfo  {journal} {Applied Physics Letters}\ }\textbf {\bibinfo {volume}
  {92}},\ \bibinfo {pages} {152505} (\bibinfo {year} {2008})}\BibitemShut
  {NoStop}%
\bibitem [{\citenamefont {Devoret}, \citenamefont {Girvin},\ and\ \citenamefont
  {Schoelkopf}(2007)}]{devoret_circuit-qed:_2007}%
  \BibitemOpen
  \bibfield  {author} {\bibinfo {author} {\bibfnamefont {M.}~\bibnamefont
  {Devoret}}, \bibinfo {author} {\bibfnamefont {S.}~\bibnamefont {Girvin}}, \
  and\ \bibinfo {author} {\bibfnamefont {R.}~\bibnamefont {Schoelkopf}},\
  }\href {\doibase 10.1002/andp.200710261} {\bibfield  {journal} {\bibinfo
  {journal} {Annalen der Physik}\ }\textbf {\bibinfo {volume} {16}},\ \bibinfo
  {pages} {767} (\bibinfo {year} {2007})}\BibitemShut {NoStop}%
\bibitem [{\citenamefont {Nigg}\ \emph {et~al.}(2012)\citenamefont {Nigg},
  \citenamefont {Paik}, \citenamefont {Vlastakis}, \citenamefont {Kirchmair},
  \citenamefont {Shankar}, \citenamefont {Frunzio}, \citenamefont {Devoret},
  \citenamefont {Schoelkopf},\ and\ \citenamefont
  {Girvin}}]{nigg_black-box_2012}%
  \BibitemOpen
  \bibfield  {author} {\bibinfo {author} {\bibfnamefont {S.~E.}\ \bibnamefont
  {Nigg}}, \bibinfo {author} {\bibfnamefont {H.}~\bibnamefont {Paik}}, \bibinfo
  {author} {\bibfnamefont {B.}~\bibnamefont {Vlastakis}}, \bibinfo {author}
  {\bibfnamefont {G.}~\bibnamefont {Kirchmair}}, \bibinfo {author}
  {\bibfnamefont {S.}~\bibnamefont {Shankar}}, \bibinfo {author} {\bibfnamefont
  {L.}~\bibnamefont {Frunzio}}, \bibinfo {author} {\bibfnamefont {M.~H.}\
  \bibnamefont {Devoret}}, \bibinfo {author} {\bibfnamefont {R.~J.}\
  \bibnamefont {Schoelkopf}}, \ and\ \bibinfo {author} {\bibfnamefont {S.~M.}\
  \bibnamefont {Girvin}},\ }\href {\doibase 10.1103/PhysRevLett.108.240502}
  {\bibfield  {journal} {\bibinfo  {journal} {Physical Review Letters}\
  }\textbf {\bibinfo {volume} {108}},\ \bibinfo {pages} {240502} (\bibinfo
  {year} {2012})}\BibitemShut {NoStop}%
\bibitem [{\citenamefont {Wallraff}\ \emph {et~al.}(2004)\citenamefont
  {Wallraff}, \citenamefont {Schuster}, \citenamefont {Blais}, \citenamefont
  {Frunzio}, \citenamefont {Huang}, \citenamefont {Majer}, \citenamefont
  {Kumar}, \citenamefont {Girvin},\ and\ \citenamefont
  {Schoelkopf}}]{wallraff_strong_2004}%
  \BibitemOpen
  \bibfield  {author} {\bibinfo {author} {\bibfnamefont {A.}~\bibnamefont
  {Wallraff}}, \bibinfo {author} {\bibfnamefont {D.~I.}\ \bibnamefont
  {Schuster}}, \bibinfo {author} {\bibfnamefont {A.}~\bibnamefont {Blais}},
  \bibinfo {author} {\bibfnamefont {L.}~\bibnamefont {Frunzio}}, \bibinfo
  {author} {\bibfnamefont {R.-S.}\ \bibnamefont {Huang}}, \bibinfo {author}
  {\bibfnamefont {J.}~\bibnamefont {Majer}}, \bibinfo {author} {\bibfnamefont
  {S.}~\bibnamefont {Kumar}}, \bibinfo {author} {\bibfnamefont {S.~M.}\
  \bibnamefont {Girvin}}, \ and\ \bibinfo {author} {\bibfnamefont {R.~J.}\
  \bibnamefont {Schoelkopf}},\ }\href {\doibase 10.1038/nature02851} {\bibfield
   {journal} {\bibinfo  {journal} {Nature}\ }\textbf {\bibinfo {volume}
  {431}},\ \bibinfo {pages} {162} (\bibinfo {year} {2004})}\BibitemShut
  {NoStop}%
\bibitem [{\citenamefont {Reed}\ \emph {et~al.}(2010)\citenamefont {Reed},
  \citenamefont {Johnson}, \citenamefont {Houck}, \citenamefont {DiCarlo},
  \citenamefont {Chow}, \citenamefont {Schuster}, \citenamefont {Frunzio},\
  and\ \citenamefont {Schoelkopf}}]{reed_fast_2010}%
  \BibitemOpen
  \bibfield  {author} {\bibinfo {author} {\bibfnamefont {M.~D.}\ \bibnamefont
  {Reed}}, \bibinfo {author} {\bibfnamefont {B.~R.}\ \bibnamefont {Johnson}},
  \bibinfo {author} {\bibfnamefont {A.~A.}\ \bibnamefont {Houck}}, \bibinfo
  {author} {\bibfnamefont {L.}~\bibnamefont {DiCarlo}}, \bibinfo {author}
  {\bibfnamefont {J.~M.}\ \bibnamefont {Chow}}, \bibinfo {author}
  {\bibfnamefont {D.~I.}\ \bibnamefont {Schuster}}, \bibinfo {author}
  {\bibfnamefont {L.}~\bibnamefont {Frunzio}}, \ and\ \bibinfo {author}
  {\bibfnamefont {R.~J.}\ \bibnamefont {Schoelkopf}},\ }\href {\doibase
  10.1063/1.3435463} {\bibfield  {journal} {\bibinfo  {journal} {Applied
  Physics Letters}\ }\textbf {\bibinfo {volume} {96}},\ \bibinfo {pages}
  {203110} (\bibinfo {year} {2010})}\BibitemShut {NoStop}%
\bibitem [{\citenamefont {Jeffrey}\ \emph {et~al.}(2014)\citenamefont
  {Jeffrey}, \citenamefont {Sank}, \citenamefont {Mutus}, \citenamefont
  {White}, \citenamefont {Kelly}, \citenamefont {Barends}, \citenamefont
  {Chen}, \citenamefont {Chen}, \citenamefont {Chiaro}, \citenamefont
  {Dunsworth}, \citenamefont {Megrant}, \citenamefont {O’Malley},
  \citenamefont {Neill}, \citenamefont {Roushan}, \citenamefont {Vainsencher},
  \citenamefont {Wenner}, \citenamefont {Cleland},\ and\ \citenamefont
  {Martinis}}]{jeffrey_fast_2014}%
  \BibitemOpen
  \bibfield  {author} {\bibinfo {author} {\bibfnamefont {E.}~\bibnamefont
  {Jeffrey}}, \bibinfo {author} {\bibfnamefont {D.}~\bibnamefont {Sank}},
  \bibinfo {author} {\bibfnamefont {J.}~\bibnamefont {Mutus}}, \bibinfo
  {author} {\bibfnamefont {T.}~\bibnamefont {White}}, \bibinfo {author}
  {\bibfnamefont {J.}~\bibnamefont {Kelly}}, \bibinfo {author} {\bibfnamefont
  {R.}~\bibnamefont {Barends}}, \bibinfo {author} {\bibfnamefont
  {Y.}~\bibnamefont {Chen}}, \bibinfo {author} {\bibfnamefont {Z.}~\bibnamefont
  {Chen}}, \bibinfo {author} {\bibfnamefont {B.}~\bibnamefont {Chiaro}},
  \bibinfo {author} {\bibfnamefont {A.}~\bibnamefont {Dunsworth}}, \bibinfo
  {author} {\bibfnamefont {A.}~\bibnamefont {Megrant}}, \bibinfo {author}
  {\bibfnamefont {P.}~\bibnamefont {O’Malley}}, \bibinfo {author}
  {\bibfnamefont {C.}~\bibnamefont {Neill}}, \bibinfo {author} {\bibfnamefont
  {P.}~\bibnamefont {Roushan}}, \bibinfo {author} {\bibfnamefont
  {A.}~\bibnamefont {Vainsencher}}, \bibinfo {author} {\bibfnamefont
  {J.}~\bibnamefont {Wenner}}, \bibinfo {author} {\bibfnamefont
  {A.}~\bibnamefont {Cleland}}, \ and\ \bibinfo {author} {\bibfnamefont
  {J.~M.}\ \bibnamefont {Martinis}},\ }\href {\doibase
  10.1103/PhysRevLett.112.190504} {\bibfield  {journal} {\bibinfo  {journal}
  {Physical Review Letters}\ }\textbf {\bibinfo {volume} {112}},\ \bibinfo
  {pages} {190504} (\bibinfo {year} {2014})}\BibitemShut {NoStop}%
\bibitem [{\citenamefont {Chou}\ \emph {et~al.}()\citenamefont {Chou},
  \citenamefont {Blumoff}, \citenamefont {Reagor}, \citenamefont {Axline},
  \citenamefont {Brierley}, \citenamefont {Nigg}, \citenamefont {Reinhold},
  \citenamefont {Heeres}, \citenamefont {Wang}, \citenamefont {Sliwa},
  \citenamefont {Narla}, \citenamefont {Hatridge}, \citenamefont {Jiang},
  \citenamefont {Devoret}, \citenamefont {Girvin},\ and\ \citenamefont
  {Schoelkopf}}]{chou_implementing_????}%
  \BibitemOpen
  \bibfield  {author} {\bibinfo {author} {\bibfnamefont {K.}~\bibnamefont
  {Chou}}, \bibinfo {author} {\bibfnamefont {J.}~\bibnamefont {Blumoff}},
  \bibinfo {author} {\bibfnamefont {M.}~\bibnamefont {Reagor}}, \bibinfo
  {author} {\bibfnamefont {C.}~\bibnamefont {Axline}}, \bibinfo {author}
  {\bibfnamefont {R.}~\bibnamefont {Brierley}}, \bibinfo {author}
  {\bibfnamefont {S.}~\bibnamefont {Nigg}}, \bibinfo {author} {\bibfnamefont
  {P.}~\bibnamefont {Reinhold}}, \bibinfo {author} {\bibfnamefont {R.~W.}\
  \bibnamefont {Heeres}}, \bibinfo {author} {\bibfnamefont {C.}~\bibnamefont
  {Wang}}, \bibinfo {author} {\bibfnamefont {K.}~\bibnamefont {Sliwa}},
  \bibinfo {author} {\bibfnamefont {A.}~\bibnamefont {Narla}}, \bibinfo
  {author} {\bibfnamefont {M.~J.}\ \bibnamefont {Hatridge}}, \bibinfo {author}
  {\bibfnamefont {L.}~\bibnamefont {Jiang}}, \bibinfo {author} {\bibfnamefont
  {M.~H.}\ \bibnamefont {Devoret}}, \bibinfo {author} {\bibfnamefont {S.~M.}\
  \bibnamefont {Girvin}}, \ and\ \bibinfo {author} {\bibfnamefont {R.~J.}\
  \bibnamefont {Schoelkopf}},\ }\href@noop {} {\ }\bibinfo {note} {(in
  preparation)}\BibitemShut {NoStop}%
\bibitem [{\citenamefont {Wang}\ \emph {et~al.}()\citenamefont {Wang},
  \citenamefont {Gao}, \citenamefont {Reinhold}, \citenamefont {Heeres},
  \citenamefont {Ofek}, \citenamefont {Chou}, \citenamefont {Axline},
  \citenamefont {Reagor}, \citenamefont {Blumoff}, \citenamefont {Sliwa},
  \citenamefont {Frunzio}, \citenamefont {Girvin}, \citenamefont {Jiang},
  \citenamefont {Mirrahimi}, \citenamefont {Devoret},\ and\ \citenamefont
  {Schoelkopf}}]{wang_schrodinger_????}%
  \BibitemOpen
  \bibfield  {author} {\bibinfo {author} {\bibfnamefont {C.}~\bibnamefont
  {Wang}}, \bibinfo {author} {\bibfnamefont {Y.~Y.}\ \bibnamefont {Gao}},
  \bibinfo {author} {\bibfnamefont {P.}~\bibnamefont {Reinhold}}, \bibinfo
  {author} {\bibfnamefont {R.~W.}\ \bibnamefont {Heeres}}, \bibinfo {author}
  {\bibfnamefont {N.}~\bibnamefont {Ofek}}, \bibinfo {author} {\bibfnamefont
  {K.}~\bibnamefont {Chou}}, \bibinfo {author} {\bibfnamefont {C.}~\bibnamefont
  {Axline}}, \bibinfo {author} {\bibfnamefont {M.}~\bibnamefont {Reagor}},
  \bibinfo {author} {\bibfnamefont {J.}~\bibnamefont {Blumoff}}, \bibinfo
  {author} {\bibfnamefont {K.~M.}\ \bibnamefont {Sliwa}}, \bibinfo {author}
  {\bibfnamefont {L.}~\bibnamefont {Frunzio}}, \bibinfo {author} {\bibfnamefont
  {S.~M.}\ \bibnamefont {Girvin}}, \bibinfo {author} {\bibfnamefont
  {L.}~\bibnamefont {Jiang}}, \bibinfo {author} {\bibfnamefont
  {M.}~\bibnamefont {Mirrahimi}}, \bibinfo {author} {\bibfnamefont {M.~H.}\
  \bibnamefont {Devoret}}, \ and\ \bibinfo {author} {\bibfnamefont {R.~J.}\
  \bibnamefont {Schoelkopf}},\ }\href {http://arxiv.org/abs/1601.05505}
  {\bibfield  {journal} {\bibinfo  {journal} {Science}\ }}\bibinfo {note} {(in
  press)}\BibitemShut {NoStop}%
\bibitem [{\citenamefont {Lecocq}\ \emph {et~al.}(2011)\citenamefont {Lecocq},
  \citenamefont {Pop}, \citenamefont {Peng}, \citenamefont {Matei},
  \citenamefont {Crozes}, \citenamefont {Fournier}, \citenamefont {{Cécile
  Naud}}, \citenamefont {Guichard},\ and\ \citenamefont
  {Buisson}}]{lecocq_junction_2011}%
  \BibitemOpen
  \bibfield  {author} {\bibinfo {author} {\bibfnamefont {F.}~\bibnamefont
  {Lecocq}}, \bibinfo {author} {\bibfnamefont {I.~M.}\ \bibnamefont {Pop}},
  \bibinfo {author} {\bibfnamefont {Z.}~\bibnamefont {Peng}}, \bibinfo {author}
  {\bibfnamefont {I.}~\bibnamefont {Matei}}, \bibinfo {author} {\bibfnamefont
  {T.}~\bibnamefont {Crozes}}, \bibinfo {author} {\bibfnamefont
  {T.}~\bibnamefont {Fournier}}, \bibinfo {author} {\bibnamefont {{Cécile
  Naud}}}, \bibinfo {author} {\bibfnamefont {W.}~\bibnamefont {Guichard}}, \
  and\ \bibinfo {author} {\bibfnamefont {O.}~\bibnamefont {Buisson}},\ }\href
  {\doibase 10.1088/0957-4484/22/31/315302} {\bibfield  {journal} {\bibinfo
  {journal} {Nanotechnology}\ }\textbf {\bibinfo {volume} {22}},\ \bibinfo
  {pages} {315302} (\bibinfo {year} {2011})}\BibitemShut {NoStop}%
\bibitem [{\citenamefont {Reed}(2013)}]{reed_entanglement_2013}%
  \BibitemOpen
  \bibfield  {author} {\bibinfo {author} {\bibfnamefont {M.}~\bibnamefont
  {Reed}},\ }\emph {\bibinfo {title} {Entanglement and {Quantum} {Error}
  {Correction} with {Superconducting} {Qubits}}},\ \href
  {http://arxiv.org/abs/1311.6759} {\bibinfo {type} {Ph.{D}. {Thesis}}},\
  \bibinfo  {school} {Yale University} (\bibinfo {year} {2013})\BibitemShut
  {NoStop}%
\bibitem [{\citenamefont {Wang}\ \emph {et~al.}(2015)\citenamefont {Wang},
  \citenamefont {Axline}, \citenamefont {Gao}, \citenamefont {Brecht},
  \citenamefont {Chu}, \citenamefont {Frunzio}, \citenamefont {Devoret},\ and\
  \citenamefont {Schoelkopf}}]{wang_surface_2015}%
  \BibitemOpen
  \bibfield  {author} {\bibinfo {author} {\bibfnamefont {C.}~\bibnamefont
  {Wang}}, \bibinfo {author} {\bibfnamefont {C.}~\bibnamefont {Axline}},
  \bibinfo {author} {\bibfnamefont {Y.~Y.}\ \bibnamefont {Gao}}, \bibinfo
  {author} {\bibfnamefont {T.}~\bibnamefont {Brecht}}, \bibinfo {author}
  {\bibfnamefont {Y.}~\bibnamefont {Chu}}, \bibinfo {author} {\bibfnamefont
  {L.}~\bibnamefont {Frunzio}}, \bibinfo {author} {\bibfnamefont {M.~H.}\
  \bibnamefont {Devoret}}, \ and\ \bibinfo {author} {\bibfnamefont {R.~J.}\
  \bibnamefont {Schoelkopf}},\ }\href {\doibase 10.1063/1.4934486} {\bibfield
  {journal} {\bibinfo  {journal} {Applied Physics Letters}\ }\textbf {\bibinfo
  {volume} {107}},\ \bibinfo {pages} {162601} (\bibinfo {year}
  {2015})}\BibitemShut {NoStop}%
\bibitem [{\citenamefont {Sandberg}\ \emph {et~al.}(2012)\citenamefont
  {Sandberg}, \citenamefont {Vissers}, \citenamefont {Kline}, \citenamefont
  {Weides}, \citenamefont {Gao}, \citenamefont {Wisbey},\ and\ \citenamefont
  {Pappas}}]{sandberg_etch_2012}%
  \BibitemOpen
  \bibfield  {author} {\bibinfo {author} {\bibfnamefont {M.}~\bibnamefont
  {Sandberg}}, \bibinfo {author} {\bibfnamefont {M.~R.}\ \bibnamefont
  {Vissers}}, \bibinfo {author} {\bibfnamefont {J.~S.}\ \bibnamefont {Kline}},
  \bibinfo {author} {\bibfnamefont {M.}~\bibnamefont {Weides}}, \bibinfo
  {author} {\bibfnamefont {J.}~\bibnamefont {Gao}}, \bibinfo {author}
  {\bibfnamefont {D.~S.}\ \bibnamefont {Wisbey}}, \ and\ \bibinfo {author}
  {\bibfnamefont {D.~P.}\ \bibnamefont {Pappas}},\ }\href {\doibase
  10.1063/1.4729623} {\bibfield  {journal} {\bibinfo  {journal} {Applied
  Physics Letters}\ }\textbf {\bibinfo {volume} {100}},\ \bibinfo {pages}
  {262605} (\bibinfo {year} {2012})}\BibitemShut {NoStop}%
\bibitem [{\citenamefont {Rigetti}\ \emph {et~al.}(2012)\citenamefont
  {Rigetti}, \citenamefont {Gambetta}, \citenamefont {Poletto}, \citenamefont
  {Plourde}, \citenamefont {Chow}, \citenamefont {Córcoles}, \citenamefont
  {Smolin}, \citenamefont {Merkel}, \citenamefont {Rozen}, \citenamefont
  {Keefe}, \citenamefont {Rothwell}, \citenamefont {Ketchen},\ and\
  \citenamefont {Steffen}}]{rigetti_superconducting_2012}%
  \BibitemOpen
  \bibfield  {author} {\bibinfo {author} {\bibfnamefont {C.}~\bibnamefont
  {Rigetti}}, \bibinfo {author} {\bibfnamefont {J.~M.}\ \bibnamefont
  {Gambetta}}, \bibinfo {author} {\bibfnamefont {S.}~\bibnamefont {Poletto}},
  \bibinfo {author} {\bibfnamefont {B.~L.~T.}\ \bibnamefont {Plourde}},
  \bibinfo {author} {\bibfnamefont {J.~M.}\ \bibnamefont {Chow}}, \bibinfo
  {author} {\bibfnamefont {A.~D.}\ \bibnamefont {Córcoles}}, \bibinfo {author}
  {\bibfnamefont {J.~A.}\ \bibnamefont {Smolin}}, \bibinfo {author}
  {\bibfnamefont {S.~T.}\ \bibnamefont {Merkel}}, \bibinfo {author}
  {\bibfnamefont {J.~R.}\ \bibnamefont {Rozen}}, \bibinfo {author}
  {\bibfnamefont {G.~A.}\ \bibnamefont {Keefe}}, \bibinfo {author}
  {\bibfnamefont {M.~B.}\ \bibnamefont {Rothwell}}, \bibinfo {author}
  {\bibfnamefont {M.~B.}\ \bibnamefont {Ketchen}}, \ and\ \bibinfo {author}
  {\bibfnamefont {M.}~\bibnamefont {Steffen}},\ }\href {\doibase
  10.1103/PhysRevB.86.100506} {\bibfield  {journal} {\bibinfo  {journal}
  {Physical Review B}\ }\textbf {\bibinfo {volume} {86}},\ \bibinfo {pages}
  {100506} (\bibinfo {year} {2012})}\BibitemShut {NoStop}%
\bibitem [{\citenamefont {Creedon}\ \emph {et~al.}(2011)\citenamefont
  {Creedon}, \citenamefont {Reshitnyk}, \citenamefont {Farr}, \citenamefont
  {Martinis}, \citenamefont {Duty},\ and\ \citenamefont
  {Tobar}}]{creedon_high_2011}%
  \BibitemOpen
  \bibfield  {author} {\bibinfo {author} {\bibfnamefont {D.~L.}\ \bibnamefont
  {Creedon}}, \bibinfo {author} {\bibfnamefont {Y.}~\bibnamefont {Reshitnyk}},
  \bibinfo {author} {\bibfnamefont {W.}~\bibnamefont {Farr}}, \bibinfo {author}
  {\bibfnamefont {J.~M.}\ \bibnamefont {Martinis}}, \bibinfo {author}
  {\bibfnamefont {T.~L.}\ \bibnamefont {Duty}}, \ and\ \bibinfo {author}
  {\bibfnamefont {M.~E.}\ \bibnamefont {Tobar}},\ }\href {\doibase
  10.1063/1.3595942} {\bibfield  {journal} {\bibinfo  {journal} {Applied
  Physics Letters}\ }\textbf {\bibinfo {volume} {98}},\ \bibinfo {pages}
  {222903} (\bibinfo {year} {2011})}\BibitemShut {NoStop}%
\bibitem [{\citenamefont {Reagor}\ \emph {et~al.}(2013)\citenamefont {Reagor},
  \citenamefont {Paik}, \citenamefont {Catelani}, \citenamefont {Sun},
  \citenamefont {Axline}, \citenamefont {Holland}, \citenamefont {Pop},
  \citenamefont {Masluk}, \citenamefont {Brecht}, \citenamefont {Frunzio},
  \citenamefont {Devoret}, \citenamefont {Glazman},\ and\ \citenamefont
  {Schoelkopf}}]{reagor_reaching_2013}%
  \BibitemOpen
  \bibfield  {author} {\bibinfo {author} {\bibfnamefont {M.}~\bibnamefont
  {Reagor}}, \bibinfo {author} {\bibfnamefont {H.}~\bibnamefont {Paik}},
  \bibinfo {author} {\bibfnamefont {G.}~\bibnamefont {Catelani}}, \bibinfo
  {author} {\bibfnamefont {L.}~\bibnamefont {Sun}}, \bibinfo {author}
  {\bibfnamefont {C.}~\bibnamefont {Axline}}, \bibinfo {author} {\bibfnamefont
  {E.}~\bibnamefont {Holland}}, \bibinfo {author} {\bibfnamefont {I.~M.}\
  \bibnamefont {Pop}}, \bibinfo {author} {\bibfnamefont {N.~A.}\ \bibnamefont
  {Masluk}}, \bibinfo {author} {\bibfnamefont {T.}~\bibnamefont {Brecht}},
  \bibinfo {author} {\bibfnamefont {L.}~\bibnamefont {Frunzio}}, \bibinfo
  {author} {\bibfnamefont {M.~H.}\ \bibnamefont {Devoret}}, \bibinfo {author}
  {\bibfnamefont {L.}~\bibnamefont {Glazman}}, \ and\ \bibinfo {author}
  {\bibfnamefont {R.~J.}\ \bibnamefont {Schoelkopf}},\ }\href {\doibase
  10.1063/1.4807015} {\bibfield  {journal} {\bibinfo  {journal} {Applied
  Physics Letters}\ }\textbf {\bibinfo {volume} {102}},\ \bibinfo {pages}
  {192604} (\bibinfo {year} {2013})}\BibitemShut {NoStop}%
\bibitem [{\citenamefont {Minev}, \citenamefont {Pop},\ and\ \citenamefont
  {Devoret}(2013)}]{minev_planar_2013}%
  \BibitemOpen
  \bibfield  {author} {\bibinfo {author} {\bibfnamefont {Z.~K.}\ \bibnamefont
  {Minev}}, \bibinfo {author} {\bibfnamefont {I.~M.}\ \bibnamefont {Pop}}, \
  and\ \bibinfo {author} {\bibfnamefont {M.~H.}\ \bibnamefont {Devoret}},\
  }\href {\doibase 10.1063/1.4824201} {\bibfield  {journal} {\bibinfo
  {journal} {Applied Physics Letters}\ }\textbf {\bibinfo {volume} {103}},\
  \bibinfo {pages} {142604} (\bibinfo {year} {2013})}\BibitemShut {NoStop}%
\bibitem [{\citenamefont {Serniak}\ and\ \citenamefont
  {Devoret}()}]{serniak_superconducting_????}%
  \BibitemOpen
  \bibfield  {author} {\bibinfo {author} {\bibfnamefont {K.}~\bibnamefont
  {Serniak}}\ and\ \bibinfo {author} {\bibfnamefont {M.~H.}\ \bibnamefont
  {Devoret}}\ }\href@noop {} \bibinfo {note} {(in
  preparation)}\BibitemShut {NoStop}%
\end{thebibliography}

\end{document}